\documentclass[reprint,superscriptaddress,amsmath,amssymb,aps,nofootinbib,pra]{revtex4-2}
\usepackage{graphicx}
\usepackage{amsmath}
\usepackage{braket}
\usepackage{hyperref}
\usepackage{overpic}
\usepackage{DejaVuSans}
\DeclareMathOperator{\tr}{tr}

\newcommand{\pf}[1]{{\usefont{T1}{DejaVuSans-TLF}{m}{n}#1}}
\begin{document}

\title{Universal Quantum Error Mitigation via Random Inverse Depolarizing Approximation}

\author{Alexander X. Miller}

\noaffiliation

\author{Micheline B. Soley}
\email{Corresponding Author: msoley@wisc.edu}

\affiliation{Department of Chemistry, University of Wisconsin-Madison, 1101 University
Ave., Madison, WI 53706, USA}

\affiliation{Department of Physics, University of Wisconsin-Madison, 1150 University
Ave., Madison, WI 53706, USA}

\affiliation{Data Science Institute, University of Wisconsin-Madison, \\ 447 Lorch Ct., Madison, WI 53706, USA}

\begin{abstract}
Given the severity of noise in near-term quantum computing, error mitigation is essential to reduce error in quantum-computer-generated expectation values. We introduce RIDA (Random Inverse Depolarizing Approximation), a simple universal method that harnesses randomly generated circuits to estimate a given circuit’s global depolarization probability and corresponding error-free expectation value. Numerical tests indicate RIDA outperforms key benchmarks, suggestive of significant accuracy improvements for applications of quantum computing across fields including physics and chemistry.
\end{abstract}

\maketitle

\textit{Introduction.} Presently, one of the key problems facing quantum information science is how to harness its potential in the face of the high error rates encountered by today's Noisy Intermediate-Scale Quantum (NISQ) computers. Many renowned quantum algorithms \cite{10.1145/237814.237866,doi:10.1137/S0097539795293172,katabarwa2024early} require large, fault-tolerant quantum computers, which exceed the scope of today's error correction algorithms \cite{campbell2017roads, harper2019fault, gottesman2022opportunitieschallengesfaulttolerantquantum}; and, although NISQ algorithms \cite{farhi2014quantumapproximateoptimizationalgorithm, Peruzzo2014, RevModPhys.94.015004, Tilly_2022, Blekos_2024} have been designed that circumvent the need for fault-tolerant quantum error correction, such algorithms nonetheless entail inherently noisy expectation values.
Advancement of quantum computing in the near-term therefore calls for error mitigation methods that use classical post-processing to estimate error-free expectation values from their noisy counterparts \cite{RevModPhys.95.045005, zimborás2025mythsquantumcomputationfault} -- ideally in a way that will be generally applicable, will involve minimal overhead, will scale well in the face of high error levels, will not require intricate noise model data subject to daily fluctuations \cite{PRXQuantum.4.010327, 10247922} or unmodeled error channels \cite{blumekohout2020wildcarderrorquantifyingunmodeled}, and will be simple to implement. 

A wide array of error mitigation methods exist today \cite{RevModPhys.95.045005}. For example, the popular Zero-Noise Extrapolation (ZNE) method \cite{PhysRevLett.119.180509, PhysRevX.8.031027, Kandala2019, Giurgica_Tiron_2020} measures circuits with systematically increased noise to extrapolate back to zero-noise limit results, but requires the additional overhead associated with multiple extrapolation circuits and frequently becomes unstable in the presence of high error rates. Since ZNE only mitigates gate error, ZNE also typically calls for complementary measurement error mitigation methods such as Twirled Readout Error Extinction (TREX) \cite{PhysRevA.105.032620}, which applies random Pauli gates prior to measurement to diagonalize the measurement error matrix and facilitate inversion. Another error mitigation method, Probabilistic Error Cancellation \cite{PhysRevLett.119.180509, vandenBerg2023, PhysRevA.109.062617, PhysRevA.109.012431}, has the theoretical ability to reproduce exact error-free expectation values based solely on noisy circuit data, but requires a large ensemble of different circuits and perfect knowledge of all error rates.

The global depolarizing model \cite{Nielsen_Chuang_2010} points to a lower-overhead approach. Errors on a quantum computer are frequently well approximated by such a model, given that errors 
or random operations tend to rapidly scramble quantum states into white noise \cite{PRXQuantum.3.010333, PRXQuantum.3.040329, dalzell2021randomquantumcircuitstransform}. Inversion of the depolarizing model can be used to readily approximate error-free expectation values with knowledge of only a circuit's noisy expectation value and its depolarization probability \cite{PhysRevLett.127.270502}.

The dilemma for implementation of such a depolarizing-model-based approach is how to accurately and efficiently approximate the depolarization probability. Exact calculation of the depolarization probability quickly becomes intractable when relying on comparison of unmitigated and error-free expectation values, as the computational cost of simulating error-free expectation values generally grows exponentially with quantum circuit size. Depolarizing-model-based approaches therefore rely on strategies such as simulation of classically emulable quantum circuits that consist chiefly of Clifford gates to fit a depolarizing-inspired model \cite{czarnik2021error}, execution of large numbers of circuits with randomized measurements \cite{PhysRevE.104.035309}, and/or the development of depolarization probability estimation circuits related to each target circuit of interest and for which the error-free expectation value is analytically known \cite{PhysRevLett.127.270502,PhysRevE.104.035309,choi2025quantumutilityscaleerrormitigation}. A state-of-the-art approach, the CNOT-only depolarization method \cite{PhysRevLett.127.270502}, estimates the depolarization probability using an estimation circuit made up of all of the target circuit's CNOT gates. Such an estimation circuit remains in the initial state in the absence of errors and thus is associated with a known expectation value of unity; however, it systematically underestimates the depolarization probability due to the construction of the estimation circuit and thus obliges supplemental error mitigation in the form of ZNE at the expense of increased overhead.

We introduce the Random Inverse Depolarizing Approximation (RIDA) method, which capitalizes on the depolarizing model to improve the accuracy of computed expectation values on NISQ computers by approximating the depolarization probability using as little as a single estimation circuit for an entire class of target circuits. The method subsequently amplifies the noisy expectation value by the corresponding factor to produce the approximate error-free expectation value. In simulated numerical experiments, we show that across all considered circuit sizes, error rates, and shot numbers, RIDA is more effective than both benchmark ZNE and CNOT-only depolarization. This positive comparison suggests broad possible impact in modern NISQ computing applications.

\textit{Method.} To establish RIDA, we consider quantum circuits associated with estimation of the expectation value of an observable $O$. 
According to the global depolarizing approximation, which models $n$-qubit noise channels as a combination of the error-free and maximally mixed state according to \cite{Nielsen_Chuang_2010}
\begin{equation}
    \rho_\epsilon=(1-p)\rho + p\frac{I^{\otimes n}}{2}\label{density_matrix},
\end{equation}
the noisy expectation value can be expressed in terms of the depolarization probability $p$ as 
\begin{equation}
    \langle O_\epsilon\rangle = (1-p)\langle O\rangle+\frac{p}{2^n}\tr{(O)}\label{depolarizing_model}.
\end{equation}
Given that any observable can be decomposed into a linear combination of Pauli strings, where the identity string is excluded due to its unit expectation value, the trace of any component Pauli string is zero such that the error-free expectation value $\langle O\rangle$ can be readily expressed in terms of its noisy counterpart $\langle O_\epsilon\rangle$ as \cite{PhysRevLett.127.270502,choi2025quantumutilityscaleerrormitigation}
\begin{equation}
    \langle O\rangle = \frac{\langle O_\epsilon\rangle}{1-p}.\label{true_expectation}
\end{equation}

To approximate the depolarization probability $p$, we employ the same relationship between error-free and noisy expectation values Eq.~\eqref{true_expectation} to estimate the depolarization probability $p^\prime$ for a closely related circuit. We recognize two conditions this circuit must fulfill: First, it must have a similar size and composition to the target circuit. Second, it must have an exactly known error-free expectation $\langle O^\prime \rangle$ for use in Eq.~\eqref{true_expectation}. The latter property is satisfied for any circuit that can be separated into a circuit followed by its inverse, which has a known, trivial error-free expectation value (unity). This type of circuit has been successfully used in randomized benchmarking for individual gates \cite{PhysRevA.77.012307,Arute2019,wallman2014randomized,harper2019statistical} and error-corrected circuits \cite{mills2025logicalaccreditationframeworkefficient}.

\begin{figure}
    \begin{overpic}
    [scale=0.18]{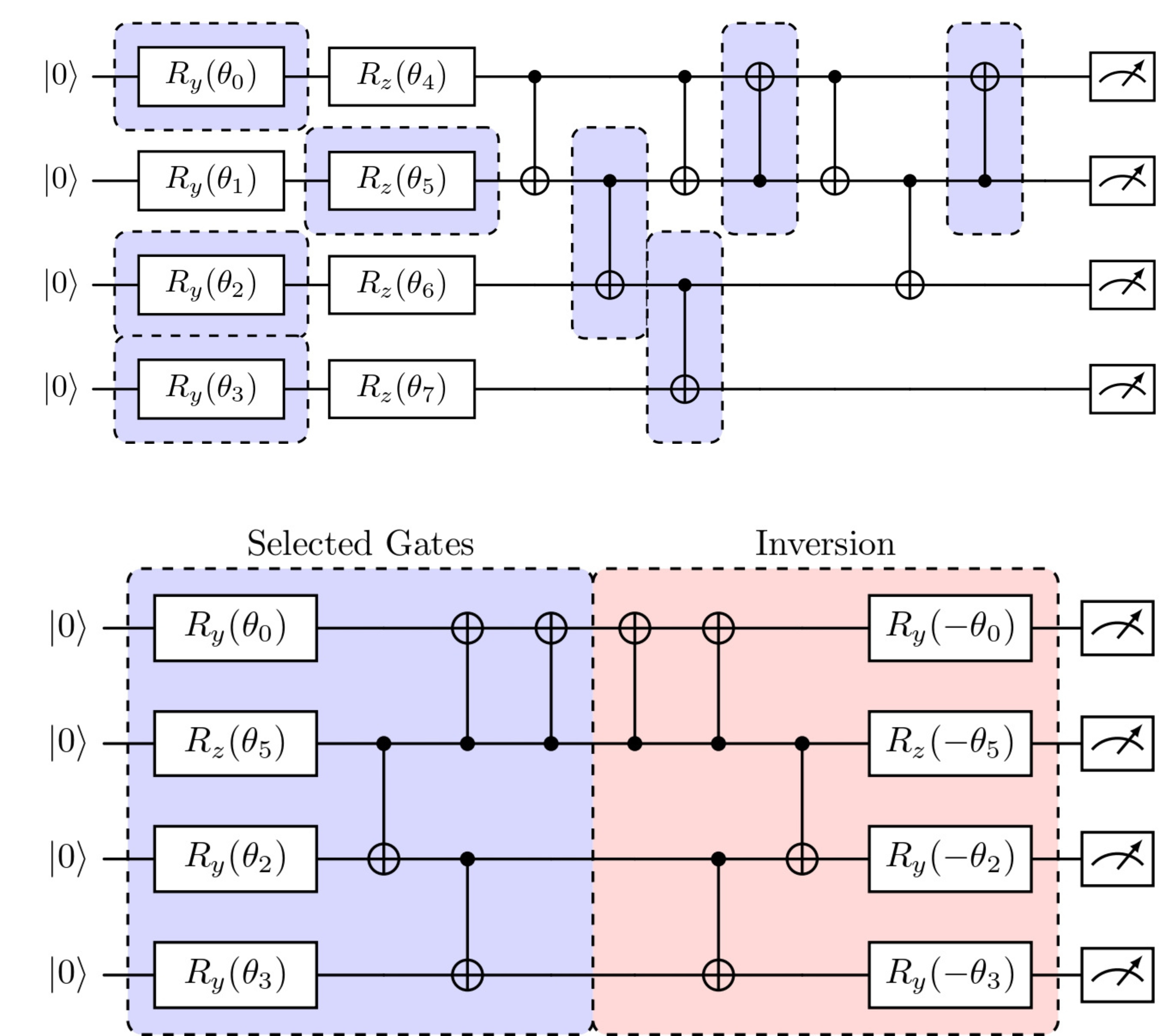}
    \put(3,88){(a)}
    \put(3,42){(b)}
    \end{overpic}
    \caption{To create a RIDA estimation circuit, (a) half of the one- and two-qubit gates in a target circuit are randomly selected (highlighted in blue) to form (b) an estimation circuit composed of the selected gates (blue) and their inverse (red).}
    \label{figure:circuit}
\end{figure}

Based on these principles, we construct straightforward RIDA estimation circuits, illustrated with an example in Fig.~\ref{figure:circuit}. We randomly select half of the gates in the target circuit to form the first half of the estimation circuit, and invert the first half to produce the second half of the estimation circuit. In order to yield the optimal distance between the depolarization probability of the estimation circuit $p^\prime$ and the true depolarization probability of the target circuit $p$, we additionally require the randomly selected gates to consist of exactly half of the one-qubit gates and half of the two-qubit gates of the target circuit (see transparent statistical argument in Sec.~I of the Supplemental Material [SM] \cite{SupplementalMaterial}).
Where a given gate is irrelevant to the measurement of $\langle O \rangle$ (\textit{i.e.,} where a gate lies on one or more ``terminal qubit(s)'' that never again interact(s) with a qubit that affects measurement of $\langle O \rangle$), we exclude the gate from the pool of possible estimation circuit gates. Additionally, where a gate straddles both one terminal and one nonterminal qubit, we add the gate to a complement circuit that is appended to the estimation circuit in order to ensure the estimation and target circuits share the same terminal qubit behavior.

\begin{figure*}
    \begin{overpic}[width=\textwidth]{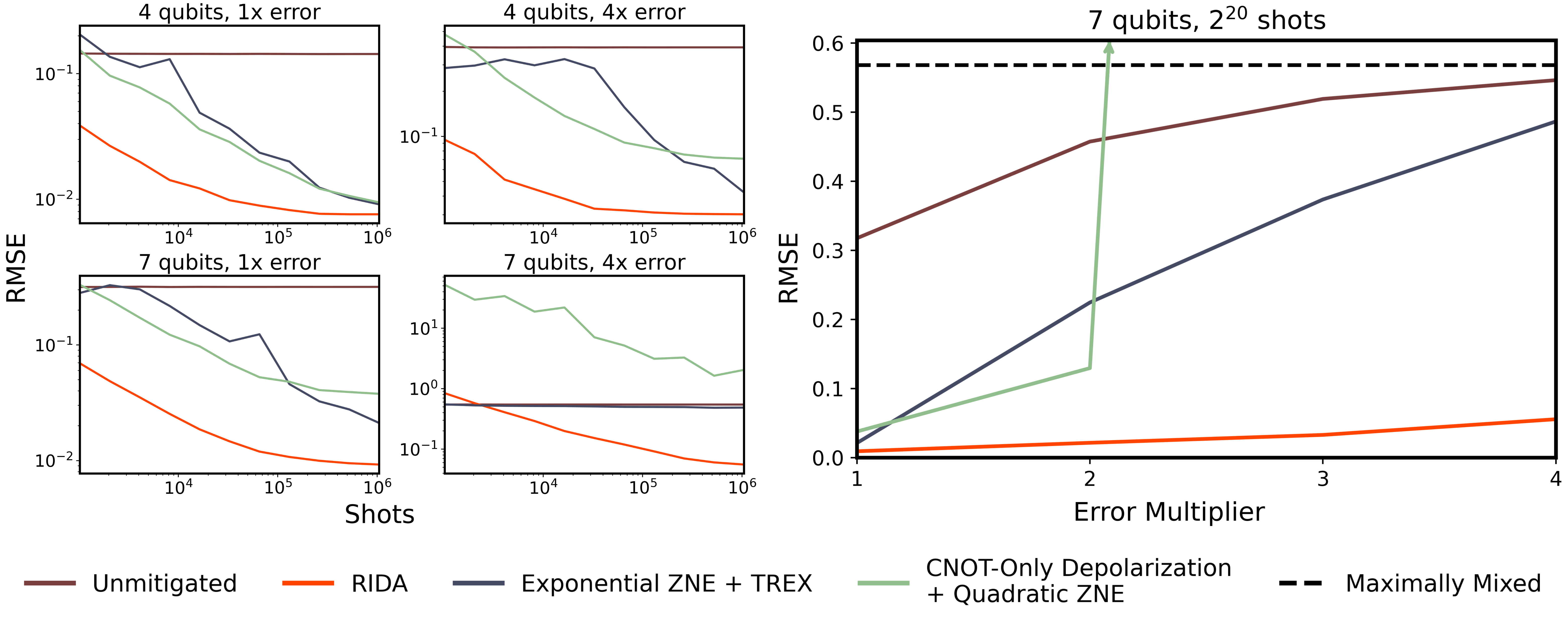}
    \put(2,38){\pf{(a)}}
    \put(25,38){\pf{(b)}}
    \put(2,22.5){\pf{(c)}}
    \put(25,23){\pf{(d)}}
    \put(51.5,38.5){\pf{(e)}}
    \end{overpic}
    \vspace*{-0.75cm}
    \caption{Root mean squared error (RMSE) for RIDA (solid orange line), exponential ZNE + TREX (solid blue line), and CNOT-only depolarization + quadratic ZNE (solid green line) as compared to unmitigated results (solid brown line) and the expectation value of the maximally mixed state (dashed black line) across a variety of qubit numbers and error multipliers. Comparisons are shown in terms of a (a)-(d) fixed error multiplier or (e) fixed shot number.}
    \label{figure:shots}
\end{figure*}

\begin{figure}
\includegraphics[scale=0.52]{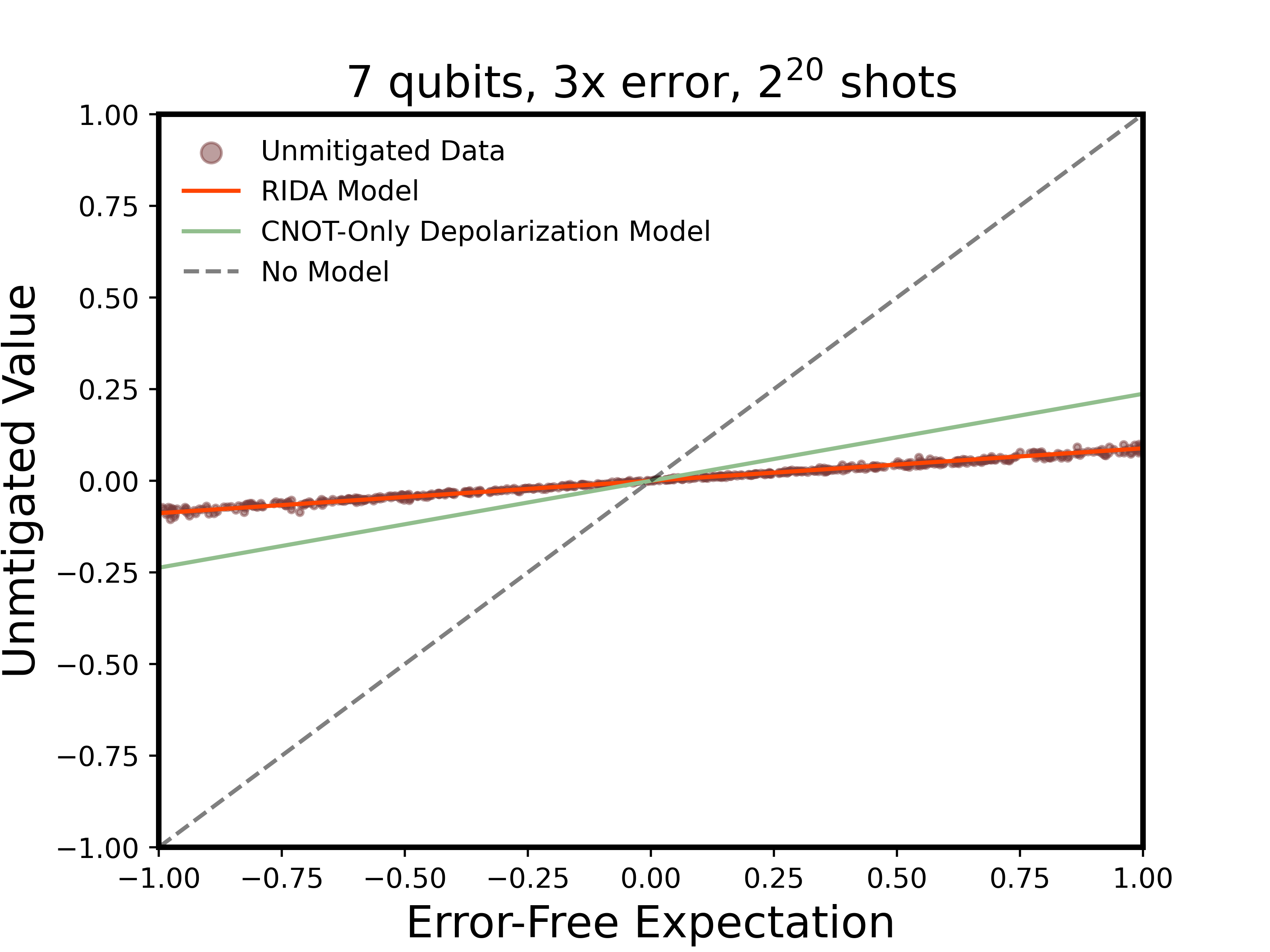}
\vspace*{-0.25cm}
\caption{Comparison of error-free and unmitigated expectation values for RIDA (orange solid line) and CNOT-only depolarization (solid green line) as compared to unmitigated data (brown circles) and the absence of a depolarizing model (dashed gray line).}
\label{figure:depolarizing}
\end{figure}

Since each gate in the target circuit has a 50\% chance of inclusion twice in the estimation circuit (once in the first half and once in inverted form in the second), the expected number and type of each gate in the target and estimation circuits are equal, which ensures comparable makeup of the two circuits. And, since for such an identity circuit $\langle O^\prime \rangle =1$ and $\tr(O^\prime)=0$, the depolarization probability of both the target and estimation circuit readily follows from the aforementioned global depolarizing model Eq.~\eqref{true_expectation}
\begin{equation}
    p\approx p^\prime=1-\langle O'_\epsilon\rangle.\label{depolarizing_estimation}
\end{equation}
Where desired, this depolarization probability may optionally be further improved via the generation of additional estimation circuits and the calculation of their average depolarization probability to reduce random variation in the computed depolarization probability.

Of note, since RIDA estimation circuits must by definition involve the same measurements as the target circuit, RIDA intrinsically mitigates measurement and gate error simultaneously. Mathematical proofs in SM Sec.~I \cite{SupplementalMaterial} indicate RIDA inherently includes measurement error mitigation that is both procedurally and mathematically equivalent to that of state-of-the-art TREX \cite{PhysRevA.105.032620} when accompanied by measurement twirling, such that RIDA requires no further measurement error mitigation.

The error-free expectation value for the target circuit then immediately follows from the relationship Eq.~\eqref{true_expectation} between the noisy expectation value from the target circuit and the depolarization probability from the estimation circuit. This is a strategy that can be reasonably used without recalculation of the depolarization probability for a range of similarly constructed target circuits with the exception of periodic updates to capture nonstationary noise fluctuations \cite{PRXQuantum.4.010327, 10247922}. We have made the RIDA code publicly available via Github \cite{ridagithub}.

\begin{figure*}
    \begin{overpic}[width=\textwidth]{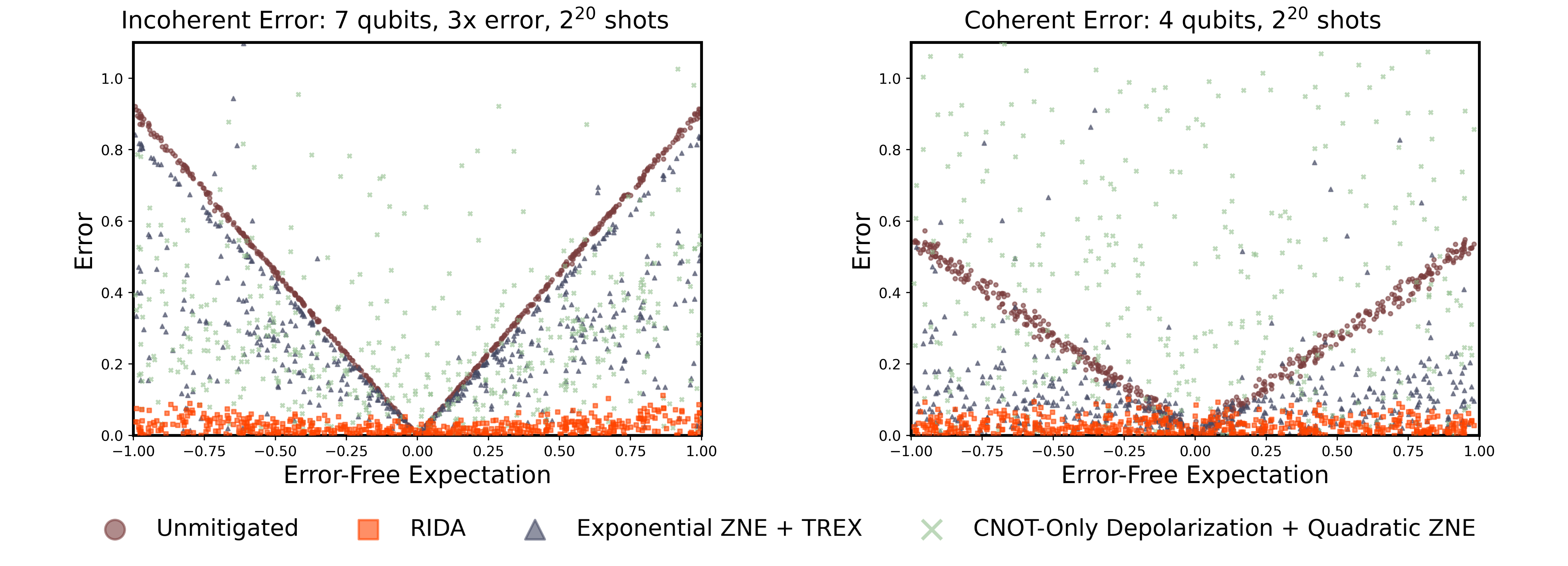}
    \put(4.5,34){\pf{(a)}}
    \put(55,34){\pf{(b)}}
    \end{overpic}
    \vspace*{-0.75cm}
    \caption{Comparison of absolute error across error-free expectation values for RIDA (orange squares), exponential ZNE + TREX (blue triangles), CNOT-only depolarization + quadratic ZNE (green crosses), and unmitigated data (brown circles) according to (a) coherent and (b) incoherent error models. Values beyond the plotted regions are truncated.}
    \label{figure:paulis}
\end{figure*}

\textit{Results.} Numerical experiments indicate RIDA outperforms benchmark methods for quantum error mitigation across combinations of qubit numbers, shot numbers, error levels, and noise models on simulated NISQ computers (see simulation details in Sec.~II of SM \cite{SupplementalMaterial}).
As seen in Fig.~\ref{figure:shots}, RIDA features a lower root mean square error (RMSE) than benchmark exponential ZNE + TREX and CNOT-only depolarization + quadratic ZNE for nearly all simulation conditions considered for an incoherent noise model based on present IBM Kingston hardware, with significant improvements over the next-best method evident from the lowest shot number and error level (75\% improvement, as shown in Fig.~\ref{figure:shots}[a]) to the highest shot number and error level (89\% improvement, as shown in Fig.~\ref{figure:shots}[d]). At the highest shot number, the error of RIDA increases at a significantly lower rate than compared methods, as exemplified by Fig.~\ref{figure:shots}(e) for seven qubits. For instance, for error multipliers in the range [1,4] (which surpass IBM Kingston error levels as a proxy for larger systems), the RMSE of RIDA spans just [0.01, 0.06] while the RMSE of exponential ZNE + TREX, CNOT-only depolarization + quadratic ZNE, and the unmitigated method span [0.02, 0.49], [0.04, 2.04], and [0.32, 0.55], respectively (with a maximally mixed state reference RMSE of 0.57). Notably, a breakdown in RIDA is only visible where shot numbers are too low and error rates too high for any error mitigation method to be practical due to universal scaling limitations \cite{PhysRevLett.131.210601}, \textit{i.e.}, where the unmitigated method itself is a poor estimate but nonetheless provides the lowest RMSE. This finding is consistent with theoretical analysis of the improvement threshold of RIDA over the unmitigated method (see SM Sec.~III \cite{SupplementalMaterial}).

The observed superior performance of RIDA over CNOT-only depolarization is also observed for the depolarization probability: As illustrated by Fig.~\ref{figure:depolarizing}, both RIDA and CNOT-only depolarization result in significantly improved estimations of the depolarization probability relative to a naive $p=0$ guess, as the resulting depolarizing model line more closely approaches the optimal fit for the unmitigated data. Whereas the relative error of the depolarization probability using 50 estimation circuits for CNOT-only depolarization is 15\%, the corresponding relative error for RIDA is just 0.2\% (or 1.0\% when using a single estimation circuit) -- a finding that supports the use of RIDA without the supplemental quadratic ZNE native to CNOT-only depolarization (an advantage that is especially important at high error levels given quadratic extrapolation only holds in the low-error limit \cite{PhysRevLett.127.270502}). 

Of key importance to practical implementation, RIDA demonstrates advantages over exponential ZNE + TREX and CNOT-only depolarization + quadratic ZNE across error-free expectation values and in the face of both incoherent and coherent errors, as shown in Fig.~\ref{figure:paulis}. In both incoherent and coherent error scenarios, the error from RIDA is low in comparison with other methods across the board, with an error below 0.11 and 0.10 for all error-free expectation values in the incoherent and coherent case, respectively. As expected, the greatest error reduction is visible at extremal error-free expectation values, where RMSE reductions of 92\% and 89\% over the next-best methods are found in the incoherent and coherent cases, respectively.

It is worthwhile to note that RIDA achieves these results with significantly lower baseline shot overhead than the alternative benchmark approaches considered since RIDA avoids the statistical variance and sampling overhead associated with execution of multiple extrapolation circuits. Statistical derivations indicate the scaling of RIDA's shot overhead coincides with the optimal result for an unbiased estimator, which represents a cubic improvement over exponential ZNE and a quintic improvement over CNOT-only depolarization + quadratic ZNE given a set of utilitarian assumptions. A simple gate-count and shot-number analysis suggests, even in the low-error limit, use of RIDA entails $16\times$ fewer shots and $47\times$ fewer gates than quadratic ZNE to achieve the same shot variance (see SM Sec.~III and extended results in Sec.~IV \cite{SupplementalMaterial}).

\textit{Conclusion.} We developed the Random Inverse Depolarizing Approximation (RIDA) method to capitalize on approximation of the global depolarization probability via randomly generated estimation circuits to consistently and efficiently mitigate quantum circuit error. The simple-to-implement RIDA method simultaneously mitigates both gate and measurement error with as little as a single estimation circuit for any target circuit of interest that is associated with estimation of an expectation value. RIDA consistently outperforms all benchmark methods considered throughout all NISQ scenarios considered, including in the presence of incoherent and coherent error. Given the observed favorable scaling of RIDA as the error level increases, RIDA is predicted to be generalizable to investigation on today's 100-qubit quantum-computing architectures.

Here we investigate RIDA for mitigation of incoherent and coherent errors, but such errors often appear in tandem with additional errors such as leakage \cite{devitt2013quantum,terhal2015quantum}, an error that plagues a wide variety of quantum computing architectures ranging from trapped-ion \cite{stricker2020experimental,moses2023race} to standard superconducting \cite{ghosh2013understanding} and dual-rail \cite{teoh2023dual,teoh2023dual1,chou2024superconducting,mehta2025bias} quantum architectures. It would therefore be valuable to determine whether RIDA can address, measure, and/or mitigate some of the difficulties imparted. Given that leakage errors share some of the same foundations that underlie RIDA, such as a similar depolarizing leakage model \cite{fowler2013coping,ghosh2013understanding,suchara2015leakage,PhysRevX.8.031027,brown2019leakage}, RIDA may prove effective in mitigating a wider range of errors beyond those discussed here.

\textit{Acknowledgments.} The authors thank Steven M. Girvin for thought-provoking discussions; the University of Wisconsin-Madison's HTCondor team and Eric H.~Wan for high-throughput computing cluster access and support, respectively; and IBM Quantum Computing for resources for quantum simulations. M.~B.~S.~acknowledges support for this research provided by the Office of the Vice Chancellor for Research and Graduate Education (OVCRGE) at the University of Wisconsin-Madison with funding from the Wisconsin Alumni Research Foundation (WARF).

\bibliography{references}

\end{document}


\title{Supplemental Material: Universal Quantum Error Mitigation via Random Inverse Depolarizing Approximation}

\author{Alexander X. Miller}

\noaffiliation

\author{Micheline B. Soley}

\affiliation{Department of Chemistry, University of Wisconsin-Madison, 1101 University
Ave., Madison, WI 53706, USA}

\affiliation{Department of Physics, University of Wisconsin-Madison, 1150 University
Ave., Madison, WI 53706, USA}

\affiliation{Data Science Institute, University of Wisconsin-Madison, \\ 447 Lorch Ct., Madison, WI 53706, USA}

\maketitle

\onecolumngrid

\tableofcontents

\section{Method Proofs and Derivations}

\subsection{Optimality of gate selection for RIDA estimation circuit generation}
\label{appendix:variance}

We show that selection of half the one-qubit gates and half the two-qubit gates from a target circuit for creation of RIDA estimation circuits (i) reduces the $p^\prime$ variance across the estimation circuits and (ii) yields optimal agreement between the average $p^\prime$ for the estimation circuits and $p$ for the target circuit, as follows:
\\

First, we establish that a fixed rather than random number of gates $G$ yields a lower $\Var (p^\prime)$. Consider that the depolarization probability $p\prime$ can be expressed as the complement of the probability of an error-free circuit execution. Where all errors are ascribed to individual gates and are considered to be independent and irreversible, the depolarization probability may be expressed in terms of the individual gate error probabilities $\{\lambda_i\}$ as
\begin{equation}
    p^\prime=1-\prod_{i=0}^G(1-\lambda_i)^2\label{error_product},
\end{equation}
where the square reflects that each gate shares the error probability of its inverse. A first-order Taylor approximation yields
\begin{equation}
    p^\prime\approx2\sum_{i=0}^G\lambda_i,\label{error_product_Taylor}
\end{equation}
which according to the rule for the variance of a constant scaling factor has a concomitant variance among estimation circuits of
\begin{equation}
    \Var(p^\prime)=4\Var\left(\sum_{i=0}^G\lambda_i\right).\label{baseline_variance}
\end{equation}

If $G$ is a constant with value $G=\E[G]$ (\textit{i.e.}, if a fixed number of gates is considered), the variance follows from that of a sum of independent random variables
\begin{equation}
    \Var(p^\prime)=4\E[G]\Var(\lambda),\label{constant_G_variance}
\end{equation}
where $\lambda$ is a random variable that describes the error rate of a randomly selected gate. However, if $G$ is instead a random variable (\textit{i.e.}, if a random number of gates were considered), the variance would follow from that of a random number of random variables
\begin{equation}
    \Var(p^\prime)=4(\E[G]\Var(\lambda)+\Var(G)(\E[\lambda])^2)\label{random_G_variance}.
\end{equation}
The variance is thus greater for a random variable $G$ than for a constant $G$, such that use of a proposed fixed number of gates is favorable.
\\

We proceed by showing that selection of a fixed number of $G_1$ one-qubit gates and $G_2$ two-qubit gates results in a lower $\Var (p^\prime)$ than selection of a random number of $G_1$ one-qubit gates and $G_2$ two-qubit gates.

If the gate numbers $G_1$ and $G_2$ are chosen randomly, we may express $\lambda_i$ in terms of random variables that represent the possible error rates of one-qubit gates $\alpha_i$ and two-qubit gates $\beta_i$ as
\begin{equation}
    \lambda_i=
    \begin{cases}
        \alpha_i\,\,\,\,\,\text{ with probability }\frac{G_1}{G}\\
        \beta_i\,\,\,\,\,\text{ with probability }\frac{G_2}{G}
    \end{cases}\label{gamma_def}
\end{equation}
and we may express $\lambda$ as
\begin{align}
    \lambda&=(1-R)\alpha+R\beta,\label{rewrite_gamma}\\
    R&=
    \begin{cases}
            0\,\,\,\,\,\text{ with probability }\frac{G_1}{G}\\
            1\,\,\,\,\,\text{ with probability }\frac{G_2}{G}
    \end{cases}.\label{Bernoulli_random_variable}
\end{align}
The law of total variance then yields
\begin{equation}
    \Var(\lambda)=\E[\Var(\lambda|R)]+\Var(\E[\lambda|R])\label{total_variance}
\end{equation}
such that the variance of $\lambda$, essential to determination of the variance of $p^\prime$, can be calculated from the expectation of the variance and the variance of the expectation.

To compute the variance of the expectation, we consider that, since $\lambda$ takes the value of either $\alpha$ or $\beta$ per Eq.~\eqref{rewrite_gamma},
\begin{equation}
    \Var(\lambda|R)=(1-R)\Var(\alpha)+R\Var(\beta)\label{var_gamma_given_R}.
\end{equation}
Application of the expression for $R$ Eq.~\eqref{Bernoulli_random_variable} to take the expectation over $R$ thus yields
\begin{equation}
    \E[\Var(\lambda|R)]=\frac{G_1}{G}\Var(\alpha)+\frac{G_2}{G}\Var(\beta).\label{total_variance_term_1}
\end{equation}

To compute the expectation of the variance, we again see that, as $\lambda$ assumes a value of $\alpha$ or $\beta$ Eq.~\eqref{rewrite_gamma}, 
\begin{equation}
    \E[\lambda|R]=(1-R)\E[\alpha]+R\E[\beta]\label{E_gamma_given_R},
\end{equation}
which may be expressed as
\begin{equation}
    \E[\lambda|R]=\E[\alpha]+R(\E[\beta]-\E[\alpha])\label{E_gamma_given_R_2},
\end{equation}
such that, since $\E[\alpha]$ is a constant and thus does not contribute to variance, the variance is
\begin{equation}
    \Var(\E[\lambda|R])=\Var((\E[\beta]-\E[\alpha])R).\label{total_variance_term_2a}
\end{equation}
Further extraction of the constant factor $\E[\beta]-\E[\alpha]$ from the variance gives
\begin{equation}
    \Var(\E[\lambda|R])=(\E[\beta]-\E[\alpha])^2\Var(R),\label{total_variance_term_2b}
\end{equation}
and substitution of the variance of $R$, which as a Bernoulli random variable Eq.~\eqref{Bernoulli_random_variable} is
\begin{equation}
    \Var(R)=\frac{G_1G_2}{G^2},\label{R_variance}
\end{equation}
gives
\begin{equation}
    \Var(\E[\lambda|R])=(\E[\beta]-\E[\alpha])^2\frac{G_1G_2}{G^2}.\label{total_variance_term_2c}
\end{equation}

The total variance of $\lambda$ where  $G_1$ and $G_2$ are chosen randomly then follows directly from substitution of the expectation of the variance Eq.~\eqref{total_variance_term_1} and the variance of the expectation Eq.~\eqref{total_variance_term_2c} into the total variance expression Eq.~\eqref{total_variance}
\begin{equation}
    \Var(\lambda)=\frac{G_1}{G}\Var(\alpha)+\frac{G_2}{G}\Var(\beta)+(\E[\beta]-\E[\alpha])^2\frac{G_1G_2}{G^2}.\label{total_variance_solved}
\end{equation}
Therefore, for constant $G$, for which the variance of $p^\prime$ in terms of $G$ Eq.~\eqref{constant_G_variance} becomes
\begin{equation}
    \Var(p^\prime)=4G\Var(\lambda),\label{two-qubit_variance_case_1a}
\end{equation}
we find
\begin{equation}
    \Var(p^\prime)=4\left(G_1\Var(\alpha)+G_2\Var(\beta)+(\E[\alpha]-\E[\beta])^2\frac{G_1G_2}{G}\right).\label{two-qubit_variance_case_1b}
\end{equation}

If the gate numbers $G_1$ and $G_2$ are instead fixed, we can instead independently add the variances associated with one- and two-qubit gates, each of which behave like the variance for a constant $G$ Eq.~\eqref{constant_G_variance}, which leads us to directly obtain the variance
\begin{equation}
    \Var(p^\prime)=4(G_1\Var(\alpha)+G_2\Var(\beta)).\label{two-qubit_variance_case_2}
\end{equation}

Since the additional term in the variance of $p^\prime$ for random $G_1,G_2$ gate numbers Eq.~\eqref{two-qubit_variance_case_1b} relative to the variance of $p^\prime$ for fixed $G_1,G_2$ gate numbers is strictly positive, the variance of $p^\prime$ is therefore lower for fixed $G_1,G_2$ gate numbers.
\\

Finally, we show that the individual gate numbers $G_1$ and $G_2$ must equal half the number of one- and two-qubit gates in the target circuit to yield the optimal (minimal) distance between the average $p^\prime$ of the estimation circuits and $p$ for the target circuit (namely, zero where the global depolarizing approximation holds). We express the Taylor approximation of the depolarization probability for the estimation circuits Eq.~\eqref{error_product_Taylor} in terms of the separate one- and two-qubit gate error contributions as
\begin{equation}
    p^\prime=2\sum_{i=0}^{G_1}\alpha_i+2\sum_{i=0}^{G_2}\beta_i\label{error_Taylor_a-b}
\end{equation}
such that the expectation value of $p^\prime$ is
\begin{equation}
    \E[p^\prime]=2G_1\E[\alpha]+2G_2\E[\beta],\label{expected_error_a-b}
\end{equation}
and we analogously express the depolarization probability $p$ for the target circuit as
\begin{equation}
    p=\sum_{i=0}^{H_1}\alpha_i+\sum_{i=0}^{H_2}\beta_i\label{original_error_Taylor_a-b}
\end{equation}
such that the expectation value of $p$ is
\begin{equation}
    \E[p]=H_1\E[\alpha]+H_2\E[\beta],\label{original_expected_error_a-b}
\end{equation}
where $H_1$ and $H_2$ are the number of one- and two-qubit gates in the target circuit, respectively.
Therefore, in order for $\E[p^\prime]$ Eq.~\eqref{expected_error_a-b} to equal $\E[p]$ Eq.~\eqref{original_expected_error_a-b}, it must be that
\begin{equation}
    2G_1\E[\alpha]+2G_2\E[\beta]=H_1\E[\alpha]+H_2\E[\beta].\label{match_expected_error}
\end{equation}
Without prior knowledge of $\E[\alpha]$ or $\E[\beta]$, this agreement is only guaranteed where
\begin{equation}
    G_1=\frac{1}{2}H_1\label{G_1}
\end{equation}
and
\begin{equation}
    G_2=\frac{1}{2}H_2,\label{G_2}
\end{equation}
and thus the choice of half of the one-qubit and half of the two-qubit gates is optimal.

\subsection{Terminal-qubit gate omission procedure}
\label{appendix:terminal}
When each RIDA estimation circuit is created, any gate in the first half of the estimation circuit is inverted in the second half of the estimation circuit prior to measurement. Thus, absent omission of certain gates from the pool of possible estimation circuit gates, a terminal qubit in the target circuit would not always result in a terminal qubit in the corresponding estimation circuit. In this scenario, an error on a terminal qubit would become reintegrated into the estimation circuit and would be included in the $p^\prime$ estimate despite the fact that such errors would have no impact on $p$. Therefore, since inclusion of such gates that only result in errors on terminal qubits would lead to overestimation of the depolarization probability, we exclude from the pool all gates that do not affect $\langle O\rangle$; namely, one-qubit gates on terminal qubits past their last point of interaction with nonterminal qubits and two-qubit gates where both qubits involved are terminal. Additionally, we append to the end of the estimation circuit a companion circuit composed of the set of all terminal-nonterminal two-qubit gates (which we decompose into controlled gates) in order to simultaneously include and exclude the errors on the nonterminal and terminal qubits, respectively.\footnote{Note there is an upper bound on the number of terminal-nonterminal gates of $n-1$ for an $n$-qubit circuit, as each qubit becomes nonterminal after it is involved in such a gate.} Since RIDA estimation circuits correspond to the identity operator, by the end of the estimation circuit the error-free state of the circuit naturally returns to the initial state $\ket{0}^{\otimes n}$, such that in the error-free state all control qubits will have reached the $\ket{0}$ state. Subsequently, since any two-qubit gate can be decomposed into a universal gate set for which the two-qubit gate in question can be represented in terms of one-qubit and controlled two-qubit gates only, the companion circuit will not alter the final state in the absence of further errors since only controlled two-qubit gates may be included in the companion circuit. In short, the estimation circuit together with the companion circuit will still correspond to an error-free expectation value of $\langle O^\prime\rangle=1$.

\subsection{Equivalence of RIDA and TREX for measurement error}
\label{appendix:measurement}

To show that RIDA measurement error mitigation with twirled readout is equivalent to TREX, we demonstrate that the measurement-error-free expectation value computed with RIDA is equivalent to that of TREX \cite{PhysRevA.105.032620} when measurement error is considered in isolation. We consider that in TREX, the estimate of the error-free expectation value is
\begin{align}
    \langle\hat{ O}_\text{TREX}\rangle&=\frac{f(D_1,s)}{f(D_0,s)},\label{TREX_expectation}\\
    f(D,s)&=\frac{1}{|D|}\sum_{(q,x)\in D}\gamma_{s,q}(-1)^{\langle s,x\rangle},\label{TREX_f}
\end{align}
where $D_1$ is the raw data and $D_0$ is the calibration data resultant from circuits executed with only measurements, each set of data $D$ contains the outputs of all measured shots in the form of a tuple of the Pauli twirling index $q$ and the measurement result $x$, and $|D|$ denotes the total shot number.
Without loss of generality, we decompose each observable into individual Pauli strings of weight $s=1$, 
such that the inner product $\langle s,x\rangle$ is the result of each shot, namely $1$ if $\ket{0}$ is measured and $-1$ if $\ket{1}$ is measured. 
The factor $\gamma_{s,q}$ then accounts for twirling applied to classical results by introducing classical bit flips where twirling would otherwise require a quantum bit flip.

Since $\ket{0}$ and $\ket{1}$ contribute $1$ and $-1$ to the calibration data function $f(D_0,s)$, respectively, the expectation value of the denominator of the TREX formula Eq.~\eqref{TREX_expectation} is equal to the noisy expectation value for the estimation circuit
\begin{equation}
    \langle f(D_0,s)\rangle=\langle O'_\epsilon\rangle;\label{TREX_f_D_0_1}
\end{equation}
and, since the error-free state contributes $1$ to $f(D_0,s)$ and the error state contributes $-1$ to $f(D_0,s)$, we obtain
\begin{align}
    \langle f(D_0,s)\rangle&=(1-p^\prime)(1)+(p^\prime)(-1)\label{TREX_f_D_0_2a}\\
    &=1-2p^\prime\label{TREX_f_D_0_2},
\end{align}
where $p^\prime$ represents the probability of measurement error on the estimation circuit.

Similarly, since $\ket{0}$ and $\ket{1}$ contribute $1$ and $-1$, respectively, to the raw data function $f(D_1,s)$, the expectation value of the denominator of the TREX formula Eq.~\eqref{TREX_expectation} is equal to the noisy expectation value corresponding to the target circuit
\begin{equation}
    \langle f(D_1, s)\rangle=\langle O_\epsilon\rangle.\label{TREX_f_D_1}
\end{equation}

Substitution of the denominator and numerator expressions Eqs.~\eqref{TREX_f_D_0_2} and \eqref{TREX_f_D_0_1} into the TREX formula Eq.~\eqref{TREX_expectation} then yields
\begin{equation}
    \langle O_\text{TREX}\rangle=\frac{\langle O_\epsilon\rangle}{1-2p^\prime},\label{TREX_expectation_2}
\end{equation}
and subsequent substitution of $p^\prime$ according to calibration data function expressions Eqs.~\eqref{TREX_f_D_0_1} and \eqref{TREX_f_D_0_2} as
\begin{equation}
    2p^\prime=1-\langle O'_\epsilon\rangle\label{TREX_p}
\end{equation}
gives
\begin{equation}
    \langle O_\text{TREX}\rangle=\frac{\langle O_\epsilon\rangle}{\langle O'_\epsilon\rangle}\label{depolarizing_no_p_2}
\end{equation}
in complete agreement with the RIDA expression
\begin{equation}
    \langle O\rangle=\frac{\langle O_\epsilon\rangle}{\langle O'_\epsilon\rangle}.\label{depolarizing_no_p}
\end{equation}
Therefore, TREX and RIDA are equivalent as long as the $\langle O'_\epsilon\rangle$ measured from TREX and RIDA are equivalent, including in the case that only measurement error mitigation is considered given that the random selection and inversion process used to generate RIDA circuits does not alter the set of qubits measured.
\\

It is worth noting that other depolarization probability estimation methods that use estimation circuits that share the same measurements and measurement-to-gate error ratio as the target circuit may also mitigate measurement error. Although not discussed in the original work \cite{PhysRevLett.127.270502}, which separately employed measurement error unfolding \cite{Nachman2020}, CNOT-only depolarization also meets the criteria needed to inherently mitigate measurement error and thereby avoid additional overhead.

\section{Simulation Details}

\subsection{Settings for numerical experiments}
\label{appendix:simulation}

Estimation circuits for RIDA, TREX, and CNOT-only depolarization are simulated using $10^7$ total shots. For RIDA and CNOT-only depolarization, which depend on nondeterministic mapping of the target circuit onto circuits with quantum computer-specific basis gates, shots are divided among 50 such circuits. Each estimation circuit is generated for a single representative of a class of sufficiently closely related target circuits (which must measure the same qubits), and is subsequently reused for other members of the class, given that such target circuits are expected to share error rate statistics. In the case of RIDA, since the depolarization probability is expected to be independent of rotation parameters, where the estimation circuit comprises only one-qubit rotations and two-qubit CNOT gates, we generate the estimation circuit for a random set of rotation parameters. 

Target circuits are simulated using the same number of total shots to compare each method. Where a given method entails multiple circuits, the total shot number is divided among the circuits run; namely, for CNOT-only depolarization and exponential Zero-Noise Extrapolation (ZNE), which both use three-point extrapolation, the shot number is divided in three (with modification to ensure the total shot number is unchanged) for each circuit corresponding to an extrapolation point.

Simulations are carried out via Qiskit with a custom incoherent error model consistent with the specifications of IBM Kingston, as described in greater detail in SM Sec.~\ref{appendix:kingston}. Where coherent noise channels are also considered, Pauli twirling (also known as randomized compiling) \cite{PhysRevA.94.052325} is employed for all methods, including twirling readout as used in TREX, to aid transformation of coherent to incoherent noise channels and thereby to facilitate use of the depolarizing approximation (see SM Sec.~\ref{appendix:coherent}).

To demonstrate the efficacy of RIDA in a wide range of experimentally relevant scenarios, we perform simulations for qubit numbers in the range $[4,7]$, multipliers of all baseline error rates in the range $[1,4]$, and target circuit shot numbers given by integer powers of two in the range $[10,20]$ (\textit{i.e.}, shot numbers of $[2^{10},2^{20}]$ or approximately one thousand to one million). Additionally, we test RIDA for a family of circuits designed to yield a uniform distribution of error-free expectation values. To produce such a distribution, for demonstration purposes, we optimize the parameters of an EfficientSU2 ansatz to produce a random target expectation value, sampled uniformly from $[-1,1]$ provided a random Pauli string with unit weight. Optimization is performed using exact expectation values computed with Qiskit statevector such that the RMSE compared to the target values is less than $10^{-5}$ to ensure the distribution does not differ appreciably from the uniform random distribution. Unless otherwise stated, all tests employ $12$ ansatz layers and consider (or average the result of) $500$ Pauli strings.

\subsection{IBM Kingston incoherent error model}
\label{appendix:kingston}

In order to simulate error comparable to that of IBM Kingston, we simulate quantum circuits in the presence of one-qubit gate, two-qubit gate, measurement, and thermal errors using the reported median error rates on the machine as of July 11, 2025 (collated in Table~\ref{tab:kingston}). Corresponding operations are applied to the density matrix after one- and two-qubit gates and after idle time for thermal errors. Gate errors specifically are modeled uniformly as local, uncorrelated depolarizing noise channels without qubit leakage. Importantly, such a simulation makes no assumptions about the validity of the global depolarizing approximation, as the local depolarizing model does not precipitate the global depolarizing model and thermal noise is inherently not depolarizing where $T_1$ and $T_2$ differ.\footnote{Here we specifically consider local depolarization to be uncorrelated; however, correlated depolarization is not expected to affect the performance of the method since multiple simultaneous errors likewise scramble into white noise consistent with the depolarizing approximation.} 

To further reflect implementation on IBM Kingston, each quantum circuit is transpiled to the basis gates available on the hardware (CZ, RX, RZ, RZZ, SX, and X).\footnote{Note that of the gates considered, RX, RZ, and RZZ are non-Clifford, such that the success of RIDA for this gate set emphasizes that the method is not limited to Clifford circuits.} Where the error mitigation methods considered normally employ CX or generic two-qubit gates, the role of such gates is instead played here by CZ, an alteration that pertains to Pauli twirling of two-qubit gates, RIDA selection of two-qubit gates, and CNOT-only depolarization. This alteration does not affect the efficacy of the methods, as the error-free output of each circuit remains unchanged, as the errors that would be associated with the simple incoherent error model considered are equivalent, and as in practice errors on CZ and CX gates are typically comparable. Additionally, each method can be readily applied to CZ gates; Pauli twirling can be applied to CZ gates by selecting a corresponding pool of operators; RIDA can be run without alteration since it is not basis-gate dependent; and CNOT-only depolarization estimation circuits can be constructed from CZ gates instead while still ensuring such circuits remain in the initial state because each control qubit still has a value of $\ket{0}$.

\begin{table}
\begin{centering}
\begin{tabular}{cc}
\toprule 
Specification & Magnitude\tabularnewline
\midrule
\midrule 
Two-qubit error & $2.07\times10^{-3}$\tabularnewline
\midrule 
One-qubit error & $2.25\times10^{-4}$\tabularnewline
\midrule 
Readout error & $7.32\times10^{-3}$\tabularnewline
\midrule 
$T_1$ (\textmu s) & $270$\tabularnewline
\midrule 
$T_2$ (\textmu s) & $143$\tabularnewline
\midrule 
Two-qubit time (\textmu s) & $6.8\times10^{-2}$\tabularnewline
\bottomrule
\end{tabular}
\par\end{centering}
\caption{Specifications for the simulations based on IBM Kingston. Where one-qubit times were not reported, a one-qubit time of one-tenth the two-qubit time was employed (to model such gates' typically lower execution time and to reflect the similar one- and two-qubit gate error rate ratio).\label{tab:kingston}}
\end{table}

\subsection{Coherent error model and twirling}
\label{appendix:coherent}
To simulate NISQ results in the face of incoherent error, we insert an $R_x$ overrotation of 0.15 radians after both qubits in each two-qubit gate. For these tests, all methods (including the unmitigated method) are supplemented by Pauli twirling the measurement. In Pauli twirling \cite{PhysRevA.77.012307}, each circuit is replaced by a number of twirled circuits, in which each two-qubit gate is sandwiched between a structure of four random Pauli gates specifically selected to ensure the outcome of noiseless execution remains unchanged (two gates per qubit, with a gate applied to each qubit both before and after the two-qubit gate). 
Twirling is also applied to measurements to facilitate inversion of the measurement error matrix, in which case the effect of certain Pauli operators is accounted for in classical post-processing \cite{PhysRevA.105.032620}. 
Results for each twirled circuit are then averaged such that the effect of noise on each gate becomes randomized over many instances to transform coherent error into incoherent error. Here simulated results for coherent tests represent averages over 250 Pauli twirled circuits total for each estimation and target circuit.

\subsection{Implementation of the CNOT-only depolarization method}
\label{appendix:cnot}

To compare RIDA to an alternative state-of-the-art depolarizing-model-based error mitigation method, we consider what we term CNOT-only depolarization \cite{PhysRevLett.127.270502}, which abstracts all of the CNOT gates from a given target circuit to form an estimation circuit. Such a construction ensures that the final state of the estimation circuit is $\ket{0}^{\otimes n}$ as long as the initial state is $\ket{0}^{\otimes n}$, given that any individual CNOT gate has null effect when applied to a $\ket{0}^{\otimes n}$ state, such that the error-free expectation value of the estimation circuit is known as required for estimation of the depolarization probability. 

However, CNOT-only depolarization is expected to systematically underestimate the depolarization probability as any one-qubit error is neglected, and, although one-qubit error is generally less than that of two-qubit error, such error is not necessarily negligible. In addition, when only CNOT-gates are considered, before any error is encountered, the estimation circuit is always in the $\ket{0}^{\otimes n}$ state, such that the state of the circuit and the interactions caused by the CNOT gates are substantially different from those occurring in the target circuit, which may feature a high degree of entanglement and qubit states anywhere on the Bloch sphere. To ameliorate this problem, the authors proposed that CNOT-only depolarization include the insertion of a layer of random rotations and its inverse at the beginning and end of the circuit, respectively. Although this formulation somewhat increases the similarity between the estimation and target circuits (and thereby is expected to improve the accuracy of the depolarization probability estimate), we find it does not always produce the desired error-free final state of $\ket{0}^{\otimes n}$ and corresponding error-free expectation value of $\langle O^\prime\rangle=1$.\footnote{For example, consider the simple two-qubit case in which the random rotation layer constitutes a bit flip on only qubit 0 and the CNOT-only block constitutes a CNOT gate where qubit 0 is the control qubit and qubit 1 is the target qubit. The state of such a system is $\ket{01}$ following application of the initial rotation; $\ket{11}$ after application of the CNOT gate; but $\ket{10}\neq\ket{00}$ after application of the inverse of the rotation.} We therefore employ CNOT-only depolarization without random rotation layers for all simulations, with the exception of the comparison of CNOT-only depolarization formulations presented in SM Sec.~\ref{appendix:rotations}.
Note the initial presentation of CNOT-only depolarization \cite{PhysRevLett.127.270502} also employed supplementary measurement error unfolding \cite{Nachman2020}; we omit this step due to the derived equivalence of measurement error mitigation in TREX and the measurement error mitigation based on the global depolarizing model with twirled readout (see SM Sec.~\ref{appendix:measurement}).
\\

Due to the underestimated depolarization probability, as in the initial presentation of CNOT-only depolarization, we consider the approach where supplemental error mitigation is provided in the form of quadratic ZNE as applied following use of the depolarizing model. In this formulation, each of the $1\times$, $3\times$, and $5\times$ error circuits correspond to separate depolarization probability estimation circuits, and are mitigated using the depolarizing model prior to quadratic ZNE. Such a procedure is expected to be superior to standard quadratic ZNE alone since initial error reduction with CNOT-only depolarization facilitates approximation of the true exponential error function as a quadratic function, as follows:

According to exponential ZNE, the error-free expectation value of a circuit may be expressed as
\begin{equation}
    \langle O\rangle=e^{a\lambda}\langle O_\epsilon\rangle,\label{exp_noise_model}
\end{equation}
where $\lambda$ is the error scaling factor and $a$ is a constant related to the noise level.
As an aside, note the exponential ZNE expression for the error-free expectation value Eq.~\eqref{exp_noise_model} is in fact equal to that of the depolarizing model 
\begin{equation}
    \langle O\rangle = \frac{\langle O_\epsilon\rangle}{1-p}\label{true_expectation}
\end{equation}
under the substitution
\begin{equation}
    e^{a}=\frac{1}{1-p}.\label{exp_depo_sub}
\end{equation}
For large $a$, this expression cannot be well approximated by a quadratic function of $\lambda$ given by the second-order Taylor approximation. However, prior reduction of the noise by some amount $b<a$ (\textit{e.g.}, via CNOT-only depolarization \cite{PhysRevLett.127.270502}) instead results in an expression
\begin{equation}
    \langle O\rangle=e^{(a-b)\lambda}\langle O_\epsilon\rangle,\label{reduced_exp_noise_model}
\end{equation}
which, by virtue of its lower exponent, is more readily approximated in terms of a quadratic function of $\lambda$. Nonetheless, for sufficiently large $a$, the exponent $a-b$ will still be too large to suit a quadratic approximation, such that CNOT-only depolarization combined with quadratic ZNE is expected to perform poorly in the presence of high levels of error.

\subsection{Implementation of quadratic ZNE}
\label{appendix:qzne}

Let $x$ be the expectation value as a function of noise strength $\lambda$. The quadratic fit for ZNE stipulates that
\begin{equation}
    x_\lambda\approx a\lambda^2+b\lambda+c,\label{quadratic_function}
\end{equation}
where $a$, $b$, and $c$ are unknown constants, such that the error-free expectation value $x_0\approx c$ may be determined upon obtaining the scaled-noise values $x_1$, $x_3$, and $x_5$ via direct solution of the system of linear equations
\begin{align}
    x_1&=a+b+c,\label{x1_qfit}\\
    x_3&=9a+3b+c,\label{x3_qfit}\\
    x_5&=25a+5b+c,\label{x5_qfit}\\
\end{align}
which yields
\begin{align}
    a &= \frac{x_5-2x_3+x_1}{8},\label{a_qfit}\\
    b &= \frac{-x_5+3x_3-2x_1}{2},\label{b_qfit}\\
    c &= \frac{3x_5-10x_3+15x_1}{8}\label{c_qfit}\\
    \,&\approx x_0.\label{x0_qfit}
\end{align}

\subsection{Implementation of exponential ZNE}
\label{appendix:ezne}

Again let $x$ be the expectation value as a function of noise strength $\lambda$. The exponential fit for ZNE assumes
\begin{equation}
    x_\lambda\approx a+be^{c\lambda},\label{exponential_function}
\end{equation}
where $a$, $b$, and $c$ are unknown constants. Acquisition of $x_1$, $x_3$, and $x_5$ allows us to consider solution of the system of equations
\begin{align}
    x_1&=a+be^c,\label{x1_efit}\\
    x_3&=a+be^{3c},\label{x3_efit}\\
    x_5&=a+be^{5c}\label{x5_efit}
\end{align}
for $a$, $b$, and $c$.
Solution of the first equation Eq.~\eqref{x1_efit} for $a$ gives
\begin{equation}
    a = x_1-be^c.\label{a_efit_1}
\end{equation}
Substitution of the expression for $a$ Eq.~\eqref{a_efit_1} into the second equation Eq.~\eqref{x3_efit} then gives
\begin{equation}
    b = \frac{x_3-x_1}{e^{3c}-e^c}.\label{b_efit_1}
\end{equation}
The value of $c$ can then be determined in two possible scenarios: $c\ne0$ and $c=0$.
\\

Where $c=0$ and $b$ becomes undefined, by Eqs.~\eqref{exponential_function}-\eqref{x5_efit},
\begin{equation}
x_1=x_3=x_5=a+b,
\end{equation}
such that it is reasonable to speculate
\begin{equation}
    x_0=x_1=x_3=x_5.\label{c=0_efit}
\end{equation}

Where $c\neq 0$, substitution of the expressions for $a$ Eq.~\eqref{a_efit_1} and $b$ Eq.~\eqref{b_efit_1} into the third equation Eq.~\eqref{x5_efit} gives an equation in $c$
\begin{align}
    x_5&=x_1-\frac{x_3-x_1}{e^{3c}-e^c}e^c+\frac{x_3-x_1}{e^{3c}-e^c}e^{5c},\label{c_efit_1}\\
    0&=(x_1-x_3)e^{5c}+(x_5-x_1)e^{3c}+(x_3-x_5)e^c.\label{c_efit_2}
\end{align}
Division by $e^c$ and use of the substitution 
\begin{equation}
    u=e^{2c}\label{u_substitution}
\end{equation} 
generates a quadratic equation in $u$
\begin{equation}
    (x_1-x_3)u^2+(x_5-x_1)u+(x_3-x_5)=0\label{u_efit_1}
\end{equation}
with solution
\begin{align}
    u&=\frac{x_1-x_5\pm \sqrt{x_5^2+x_1^2+2x_1x_5+4x_3^2-4x_1x_3-4x_3x_5}}{2(x_1-x_3)}\label{u_efit_2}\\
    &=\frac{x_1-x_5\pm \sqrt{(x_1-2x_3+x_5)^2}}{2(x_1-x_3)}\label{u_efit_3}\\
    &=\frac{x_1-x_5\pm (x_1-2x_3+x_5)}{2(x_1-x_3)}.\label{u_efit_4}
\end{align}
The solution of the quadratic equation that satisfies the requirement $c\ne0$ is 
\begin{equation}
    u=\frac{x_3-x_5}{x_1-x_3}\label{u_efit_5}
\end{equation}
such that in terms of $u$, the expressions for $c$  Eq.~\eqref{u_substitution}, $b$ Eq.~\eqref{b_efit_1}, and $a$ Eq.~\eqref{a_efit_1} are
\begin{align}
    c&=\ln(\sqrt{u}),\label{c_efit_3}\\
    b&=\frac{x_3-x_1}{\sqrt{u}(u-1)},\label{b_efit-2}\\
    a &= x_1-\frac{x_3-x_1}{u-1}.\label{a_efit_2}
\end{align}
The error-free expectation value $x_0$ for the exponential fit Eq.~\eqref{exponential_function}
\begin{equation}
    x_0=a+b\label{x0_efit_1}
\end{equation}
then follows as
\begin{equation}
    x_0=x_1+\frac{x_1-x_3}{u+\sqrt{u}}.\label{x0_efit_2}
\end{equation}
Note in order to produce a valid three-point exponential fit, this solution must be finite and real, and thus $u$ must be positive and well defined.
According to Eq.~\eqref{u_efit_5}, $u$ will not be a positive, well-defined number where 
$x_\lambda$ is not a monotonic function. There are three cases in which this occurs: where $x_1$ lies between $x_3$ and $x_5$, where $x_5$ lies between $x_1$ and $x_3$, and where $x_1=x_5\neq x_3$.
\\

We address these three anomalous exponential ZNE cases as follows:

Where $x_1$ lies between $x_3$ and $x_5$, we recognize that the optimal least-squares fit function remains nearly constant between $x_1$ and $x_3$ followed by a sharp spike that reproduces $x_5$ exactly. In this case, we use this as a reasonable zero-noise expectation value
\begin{equation}
    x_0=\frac{x_1+x_3}{2}.\label{edge_case_1_efit}
\end{equation}

Where $x_5$ lies between $x_1$ and $x_3$, the optimal least-squares fit function analogously remains nearly constant between $x_3$ and $x_5$ followed by a sharp spike that reproduces $x_1$ exactly. However, given that the function continues to rise rapidly below $x_1$, this line of reasoning produces the intractable zero-noise expectation value
\begin{equation}
    x_0\rightarrow\pm\infty.\label{edge_case_2_behavior_efit}
\end{equation}
Therefore, we forgo the optimal least-squares estimate for this case and instead employ the linear extrapolation result
\begin{equation}
    x_0=\frac{3x_1-x_3}{2}. \label{edge_case_2_efit}
\end{equation}

Lastly, where $x_1=x_5\neq x_3$, both of the least-squares fit functions of the first two cases are optimal, as the two functions yield the same least-squares error. This concurrency implies both of the least-squares solutions from the first case Eq.~\eqref{edge_case_1_efit} and the second case Eq.~\eqref{edge_case_2_behavior_efit} hold such that the solution is not well defined. Thus, we develop an alternative solution based on the finding that, in practical numerical simulations, additional noise causes monotonic decay to the maximally mixed state, and thus the non-monotonic behavior of $x_1=x_5\neq x_3$ is a result of shot error. Since the shot error is at least comparable to the difference between data points, it is not meaningful to extrapolate and we use
\begin{equation}
    x_0=x_1.\label{edge_case_3_efit}
\end{equation}
 
\section{Analytic Scaling Derivations}

\subsection{Dependence of shot variance on shot number}
\label{appendix:shots}
To determine the shot variance of the estimate of $\langle O_\epsilon\rangle$ for the purpose of the following analytic scaling derivations, we consider that, where $\langle O_\epsilon\rangle$ falls in the range $[-1,1]$, it can be represented as
\begin{equation}
    \langle O_\epsilon\rangle=1-2r,\label{o-r}
\end{equation}
where $r$ is a classical probability of measuring some output state. Given that $r$ is a classical probability, the estimate $\hat{r}$ must follow a binomial distribution, such that its variance is
\begin{equation}
    \Var(\hat{r})=\frac{r(1-r)}{s},\label{var_r1}
\end{equation}
where $s$ is the number of shots. Consideration of the relationship between $\langle O_\epsilon\rangle$ and $r$ Eq.~\eqref{o-r} and the variance of a constant multiple of a variable then yields
\begin{equation}
    \Var(\hat{\langle O_\epsilon\rangle})=4\Var(\hat{r}).\label{var_o-r}
\end{equation}
Thus, substitution of the expression for $r$ in terms of $\langle O_\epsilon\rangle$ Eq.~\eqref{o-r} into the variance of $\hat{r}$ in terms of $r$ Eq.~\eqref{var_r1} gives
\begin{equation}
    4\Var(\hat{r})=\frac{1-\langle O_\epsilon\rangle^2}{s},\label{var_r2}
\end{equation}
and substitution of the resulting expression into the relationship between the variances of $\hat{r}$ and $\hat{\langle O_\epsilon\rangle}$ Eq.~\eqref{var_o-r} gives the shot variance of $\hat{\langle O_\epsilon\rangle}$
\begin{equation}
    \Var(\hat{\langle O_\epsilon\rangle})=\frac{1-\langle O_\epsilon\rangle^2}{s}.\label{var_o}
\end{equation}
Notably, in the extreme high-error limit $p\rightarrow1$, $\langle O_\epsilon\rangle\rightarrow 0$, and thus
\begin{equation}
    \Var(\hat{\langle O_\epsilon\rangle})\approx\frac{1}{s}.\label{var_o_approx}
\end{equation}
Finally, methods that perform three-point extrapolation must use $1/3$ as many shots for each point,\footnote{As the number of shots is not always divisible by three, we divide the number of shots as evenly as possible between the three circuits and allocate the remaining shot(s) to the first extrapolation circuit. This allocation ensures a negligible difference in the number of shots for each circuit from that of a uniform distribution given a sufficient number of total shots (\textit{i.e.}, a distinction of one or two shots out of on the order of one thousand or one million shots).} and thus
\begin{align}
    \Var(\hat{\langle O_\epsilon\rangle})&=\frac{1-\langle O_\epsilon\rangle^2}{s/3}\\
    &=\frac{3-3\langle O_\epsilon\rangle^2}{s},\label{var_o_extrapolate}
\end{align}
and, in the high-error limit,
\begin{align}
    \Var(\hat{\langle O_\epsilon\rangle})&\approx\frac{3}{s}.\label{var_o_extrapolate_approx}
\end{align}

\subsection{Improvement threshold of RIDA}
\label{appendix:cutoff}

To alleviate the possible concern that RIDA amplifies the shot error in the process of amplifying the noisy expectation value to estimate the error-free expectation value Eq.~\eqref{true_expectation}, we analytically derive the cutoff depolarization probability and shot number where such an amplification would cause the RMSE of the RIDA estimate to surpass that of the unmitigated result under the assumption that the depolarizing approximation is accurate.

We first consider the mean squared error of the unmitigated expectation value. We can use the definition of variance to decompose this mean squared error into the difference between noisy and error-free expectations plus the shot variance
\begin{equation}
    \E[(\langle O\rangle-\hat{\langle O_\epsilon\rangle})^2]=(\E[\langle O\rangle-\hat{\langle O_\epsilon\rangle}])^2+\Var(\hat{\langle O_\epsilon\rangle}).\label{unmitigated_MSE_setup}
\end{equation}
We can evaluate the difference in expectations as
\begin{equation}
    (\E[\langle O\rangle-\hat{\langle O_\epsilon\rangle}])^2=(\langle O\rangle-\langle O_\epsilon\rangle)^2\label{nonshot_MSE_unmitigated}
\end{equation}
and use the shot variance Eq.~\eqref{var_o} derived in SM. Sec~\ref{appendix:shots}
\begin{equation}
    \Var(\hat{\langle O_\epsilon\rangle})=\frac{1-\langle O_\epsilon\rangle^2}{s};\label{shot_MSE_unmitigated_2}
\end{equation}
thus we define the total mean squared error $\mathcal{E}_\text{raw}$ to be
\begin{equation}
    \E[(\langle O\rangle-\hat{\langle O_\epsilon\rangle})^2]=\mathcal{E}_\text{raw}=(\langle O\rangle-\langle O_\epsilon\rangle)^2+\frac{1-\langle O_\epsilon\rangle^2}{s}.\label{MSE_unmitigated}
\end{equation}
Application of the expression for the noisy expectation value in terms of the depolarization probability Eq.~\eqref{true_expectation}
and subsequent reorganization results in
\begin{equation}
    \mathcal{E}_\text{raw}=\left((2-p)^2-\frac{(1-p)^2}{s}\right)\langle O\rangle^2+\frac{1}{s}.\label{MSE_unmitigated_2}
\end{equation}
Since $\langle O\rangle$ is not generally known in advance in practical applications, we consider the mean squared error across $\langle O\rangle$ values, which we obtain with the integral
\begin{equation}
    \bar{\mathcal{E}}_\text{raw}=\frac{1}{2}\int_{-1}^1\left(\left((2-p)^2-\frac{(1-p)^2}{s}\right)\langle O\rangle^2+\frac{1}{s}\right)\,\textrm{d}\langle O\rangle,\label{unmitigated_avg_1}
\end{equation}
which evaluates to
\begin{align}
    \bar{\mathcal{E}}_\text{raw}&=\frac{(2-p)^2}{3}-\frac{(1-p)^2}{3s}+\frac{1}{s}\label{unmitigatd_avg_2}\\
    &=\frac{2+2p-p^2+s(2-p)^2}{3s}.\label{unmitigated_avg_3}
\end{align}

We then consider the mean squared error of the error-free expectation value from RIDA. Given that RIDA amplifies the unmitigated result by a factor of $\frac{1}{1-p}$ to arrive at the RIDA estimate Eq.~\eqref{true_expectation}, we apply the rule for variance of a constant factor to find that the RIDA shot variance is a $(\frac{1}{1-p})^2$-multiple of the unmitigated shot variance Eq.~\eqref{shot_MSE_unmitigated_2}, \textit{i.e.,}
\begin{equation}
    \mathcal{E}_\text{RIDA}=\frac{1-\langle O_\epsilon\rangle^2}{s(1-p)^2}.\label{MSE_RIDA_1}
\end{equation}
Again, we employ the relationship between the noisy expectation value and the depolarization probability Eq.~\eqref{true_expectation} to obtain
\begin{equation}
    \mathcal{E}_\text{RIDA}=\frac{1-(1-p)^2\langle O\rangle^2}{s(1-p)^2},\label{MSE_RIDA_2}
\end{equation}
which we can then integrate over the possible error-free expectation values
\begin{equation}
    \bar{\mathcal{E}}_\text{RIDA}=\frac{1}{2}\int_{-1}^1\left(\frac{1-(1-p)^2\langle O\rangle^2}{s(1-p)^2}\right)\,\textrm{d}\langle O\rangle\label{RIDA_avg_1}
\end{equation}
to obtain
\begin{equation}
    \bar{\mathcal{E}}_\text{RIDA}=\frac{2+2p-p^2}{3s(1-p)^2},\label{RIDA_avg_2}
\end{equation}
the total mean squared error of the RIDA estimate.

\begin{figure}
    \includegraphics[scale=0.6]{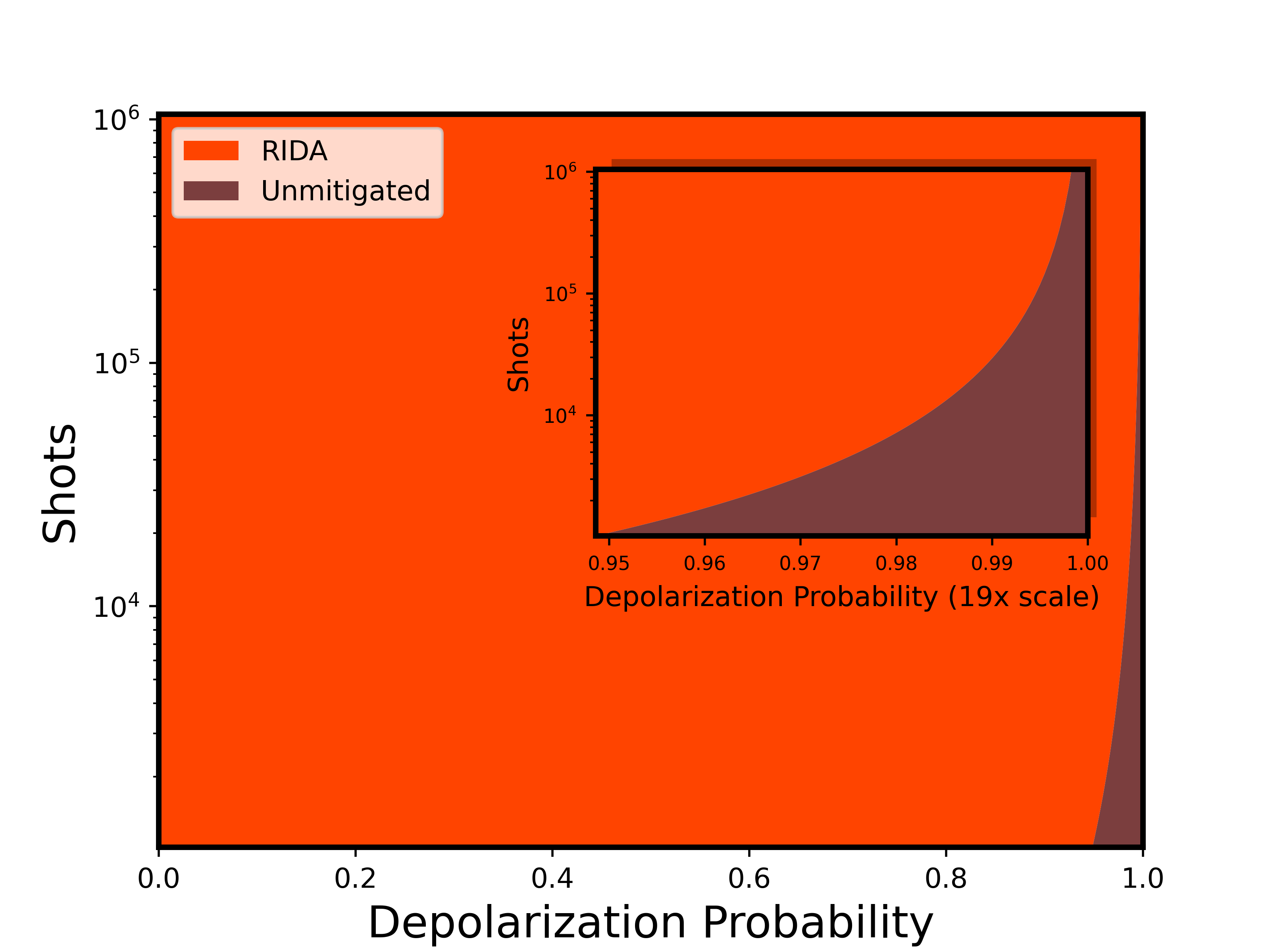}
    \caption{Analytic prediction of where the RIDA (orange) or unmitigated (brown) result has a lower RMSE as a function of depolarization probability and shot number. Inset magnifies the high-$p$ limit, \textit{i.e.}, the only cases for which RIDA's RMSE is higher than that of the unmitigated method.}
    \label{figure:cutoff}
\end{figure}

The mean squared error of RIDA Eq.~\eqref{RIDA_avg_2} is therefore less than that of the unmitigated results Eq.~\eqref{unmitigated_avg_3} where
\begin{align}
    \frac{2+2p-p^2}{3s(1-p)^2}&<\frac{2+2p-p^2+s(2-p)^2}{3s},\label{RIDA_MSE_comp_1}\\
    s&>\frac{(2+2p-p^2)}{(2-p)^2}\left(\frac{1}{(1-p)^2}-1\right),\label{RIDA_MSE_comp_2}\\
    s&>\frac{p(2+2p-p^2)}{(2-p)(1-p)^2},\label{RIDA_MSE_comp_3}
\end{align}
such that, even in the face of shot-error amplification, RIDA will outperform unmitigated results in the vast majority of cases (see Fig.~\ref{figure:cutoff}). The sole exception is in the extreme high-depolarization probability limit $p\rightarrow1$ with low shot number $s$. Interestingly, when this case arises, $\langle O_\epsilon\rangle\rightarrow0$ regardless of the error-free expectation $\langle O \rangle$ such that the only case in which the RIDA estimate is surpassed by the unmitigated result is in the case of such overwhelming noise that essentially neither approach is effective for extraction of information.

\subsection{RIDA overhead}
\label{appendix:scaling}
To determine the scaling of the sampling overhead of RIDA, we consider the variance of the RIDA estimate of the error-free expectation value Eq.~\eqref{true_expectation}, which by the formula for variance of a constant factor is
\begin{equation}
    \Var(\hat{\langle O\rangle})=\left(\frac{1}{1-p}\right)^2\Var(\hat{\langle O_\epsilon\rangle}).
\end{equation}
Substitution of the variance of $\hat{\langle O_\epsilon\rangle}$ Eq.~\eqref{var_o_approx}\footnote{Note this substitution involves an additional implicit assumption that the system is in the extreme high-error limit; below this limit, the numerator is instead $1-\langle O_\epsilon \rangle^2$, such that the shot variance is expected to be lower.} and expression of $\Var(\hat{\langle O\rangle})$ as $\sigma^2$ leads to
\begin{equation}
    \sigma^2=\frac{1}{s(1-p)^2}\label{sigma}
\end{equation}
and thus
\begin{equation}
    s=\frac{1}{\sigma^2(1-p)^2}.\label{rida_shots}
\end{equation}
Where we consider the circuit in terms of $L$ independent layers each with single-layer depolarization probability $p^*$, equating the probabilities of no errors occurring in each formulation yields
\begin{equation}
    (1-p)=(1-p^*)^L,\label{rida_pstar}
\end{equation}
such that the expression for $s$ Eq.~\eqref{rida_shots} may be expressed as
\begin{equation}
    s=\frac{1}{\sigma^2(1-p^*)^{2L}},\label{rida_shots_2}
\end{equation}
or equivalently, where the error strength \cite{PhysRevLett.131.210601} is defined as
\begin{equation}
    \gamma=\frac{1}{1-p^*},\label{gamma}
\end{equation}
as
\begin{equation}
    s=\frac{\gamma^{2L}}{\sigma^2}.\label{rida_scaling_1}
\end{equation}

To facilitate direct comparison to literature results, recall that RIDA here is presented as applied to a single Pauli string of unit weight; the result for a combination of Pauli strings with different weights follows from the variance of a linear combination of random variables as
\begin{equation}
    s=\frac{||\boldsymbol{x}||^2}{\sigma^2}\gamma^{2L},\label{rida_scaling_2}
\end{equation}
where $\boldsymbol{x}$ is a vector of weights. This scaling is the optimal shot error scaling for an error mitigation strategy that produces unbiased estimators \cite{PhysRevLett.131.210601}.

\subsection{Exponential ZNE overhead}
\label{appendix:zne_scaling}

In order to analytically identify the scaling of the sampling overhead of exponential ZNE in the same manner as RIDA (which requires expression of the shot error as a multiple of the unmitigated value), we make the following two assumptions:
\begin{enumerate}
    \item The exponential extrapolation function Eq.~\eqref{exponential_function} and depolarizing model Eq.~\eqref{true_expectation} constitute accurate representations of the error.
    \item The shot error is sufficiently low such that the error-free estimation value $x_0$ can be approximated linearly as a function of $x_1$, $x_3$, and $x_5$.
\end{enumerate}
Use of these assumptions is substantiated by numerical simulations given the agreement between the numerical mean squared error and the analytic shot variance in regions where shot error dominates (SM. Sec~\ref{appendix:predictions}).

Where these assumptions hold, given that the estimate of the error-free expectation $x_0$ Eq.~\eqref{x0_efit_2} depends on $x_1$, $x_3$, and $x_5$ implicitly through $u$ Eq.~\eqref{u_efit_5}, we linearly approximate $x_0$ as a function of $x_1$, $x_3$, and $x_5$
\begin{equation}
    x_0\approx\frac{\partial x_0}{\partial x_1}x_1+\frac{\partial x_0}{\partial x_3}x_3+\frac{\partial x_0}{\partial x_5}x_5+k,\label{x_0_approx}
\end{equation}
where $k$ is an arbitrary constant, such that according to the variance of a linear combination of random variables,
\begin{equation}
    \Var(x_0)=\left(\frac{\partial x_0}{\partial x_1}\right)^2\Var(x_1)+\left(\frac{\partial x_0}{\partial x_3}\right)^2\Var(x_3)+\left(\frac{\partial x_0}{\partial x_5}\right)^2\Var(x_5).\label{var_x0_1}
\end{equation}
Since $x_1$, $x_3$, and $x_5$ each individually have a shot variance of $3/s$ Eq.~\eqref{var_o_extrapolate_approx},\footnote{Again this approximation assumes implicitly the extreme high-error limit, such that the shot variance at lower error levels may in practice be marginally lower.} the variance becomes
\begin{align}
    \Var(x_0)&=\frac{3}{s}\left(\left(\frac{\partial x_0}{\partial x_1}\right)^2+\left(\frac{\partial x_0}{\partial x_3}\right)^2+\left(\frac{\partial x_0}{\partial x_5}\right)^2\right)\label{var_x0_2}\\
    &=\frac{||\nabla x_0||_2^2}{s}.\label{var_x0_3}
\end{align}
To evaluate $\nabla x_0$, we calculate the derivative of $x_0$ with respect to $x_1$ as
\begin{align}
    \frac{\partial x_0}{\partial x_1}&=\frac{\partial}{\partial x_1}\left(x_1+\frac{x_1-x_3}{u+\sqrt{u}}\right)\label{del_x1_1}\\
    &=1+(u+\sqrt{u})^{-1}-(x_1-x_3)(u+\sqrt{u})^{-2}\left(\frac{1}{2}u^{-1/2}+1\right)\left(\frac{\partial u}{\partial x_1}\right)\label{del_x1_2}\\
    &=1+(u+\sqrt{u})^{-1}+u(u+\sqrt{u})^{-2}\left(\frac{1}{2}u^{-1/2}+1\right),\label{del_x1_3}
\end{align}
with respect to $x_3$ as
\begin{align}
    \frac{\partial x_0}{\partial x_3}&=\frac{\partial}{\partial x_3}\left(x_1+\frac{x_1-x_3}{u+\sqrt{u}}\right)\label{del_x3_1}\\
    &=-(u+\sqrt{u})^{-1}-(x_1-x_3)(u+\sqrt{u})^{-2}(\frac{1}{2}u^{-1/2}+1)\left(\frac{\partial u}{\partial x_3}\right)\label{del_x3_2}\\
    &=-(u+\sqrt{u})^{-1}+(u+\sqrt{u})^{-2}(\frac{1}{2}u^{-1/2}+1)(u-1),\label{del_x3_3}
\end{align}
and with respect to $x_5$ as
\begin{align}
    \frac{\partial x_0}{\partial x_5}&=\frac{\partial}{\partial x_5}\left(x_1+\frac{x_1-x_3}{u+\sqrt{u}}\right)\label{del_x5_1}\\
    &=-(x_1-x_3)(u+\sqrt{u})^{-2}\left(\frac{1}{2}u^{-1/2}+1\right)\left(\frac{\partial u}{\partial x_5}\right)\label{del_x5_2}\\
    &=(u+\sqrt{u})^{-2}\left(\frac{1}{2}u^{-1/2}+1\right).\label{del_x5_3}
\end{align}
The magnitude of the gradient is then
\begin{align}
    ||\nabla x_0||_2&=
    \Biggl(
    \left(1+(u+\sqrt{u})^{-1}+u(u+\sqrt{u})^{-2}\left(\frac{1}{2}u^{-1/2}+1\right)\right)^2\Biggr.\nonumber\\
    &+\left(-(u+\sqrt{u})^{-1}+(u+\sqrt{u})^{-2}(\frac{1}{2}u^{-1/2}+1)(u-1)\right)^2\nonumber\\
    &+\Biggl.\left((u+\sqrt{u})^{-2}\left(\frac{1}{2}u^{-1/2}+1\right)\right)^2
    \Biggr)^{1/2}.\label{ezne_grad_1}
\end{align}

To express this magnitude in terms of $\gamma$ to produce the shot variance scaling, we first determine the dependence of $u$ on $\gamma$ Eq.~\eqref{gamma}, which in turn requires knowledge of the dependence of $u$ on $p$ where the depolarizing approximation holds. To find the latter dependence, we apply the depolarizing approximation Eq.~\eqref{true_expectation} first to the relationship between $x_1$ and $x_2$ (a hypothetical circuit with $2\times$ error). Where we assume the probability that an error occurs on $x_2$ but not $x_1$ is equal to the probability $p$ that an error occurs on $x_1$ but not $x_0$, we can consider $x_1$ as the corresponding ``error-free" analog of $x_2$ instead of $x_0$ to $x_1$. The value of $x_2$ then follows from the depolarizing approximation Eq.~\eqref{true_expectation} as
\begin{equation}
    x_2=(1-p)x_1\label{x2-x1_relationship},
\end{equation}
and an analogous argument for $x_3$ leads to
\begin{equation}
    x_3=(1-p)x_2=(1-p)^2x_1.\label{x3-x1_relationship}
\end{equation}
Likewise, similar logic leads to the relationship between $x_5$ and $x_3$
\begin{equation}
    x_5=(1-p)^2x_3.\label{x5-x3_relationship}
\end{equation}
Substitution of the expressions for $x_3$ and $x_5$ (Eqs.~\eqref{x3-x1_relationship} and \eqref{x5-x3_relationship}, respectively) into the expression for $u$ Eq.~\eqref{u_efit_5} then gives
\begin{align}
    u&=\frac{(1-p)^2(x_1-x_3)}{x_1-x_3}\label{u-p_relationship_1}\\
    &=(1-p)^2.\label{u-p_relationship_3}
\end{align}

To relate the resulting $u$ in terms of $\gamma$ instead of in terms of $p$, we apply the relationship between the full depolarization probability $p$ and the single-layer depolarization probability $p^*$ Eq.~\eqref{rida_pstar} to find
\begin{equation}
    u=(1-p^*)^{2L}\label{u-p'},
\end{equation}
which by the definition of the error strength $\gamma$ Eq.~\eqref{gamma} is
\begin{equation}
    u=\gamma^{-2L}.\label{u-gamma}
\end{equation}
We can then reformulate the gradient Eq.~\eqref{ezne_grad_1} in terms of $\gamma$
\begin{align}
    ||\nabla x_0||_2&=\Biggl(
    \left(1+(\gamma^{-2L}+\gamma^{-L})^{-1}+\gamma^{-2L}(\gamma^{-2L}+\gamma^{-L})^{-2}\left(\frac{1}{2}\gamma^{L}+1\right)\right)^2\Biggr.\nonumber\\
    &+\left(-(\gamma^{-2L}+\gamma^{-L})^{-1}+(\gamma^{-2L}+\gamma^{-L})^{-2}(\frac{1}{2}\gamma^{L}+1)(\gamma^{-2L}-1)\right)^2\nonumber\\
    &+\Biggl.\left((\gamma^{-2L}+\gamma^{-L})^{-2}\left(\frac{1}{2}\gamma^{L}+1\right)\right)^2
    \Biggr)^{1/2}\label{ezne_grad_2}
\end{align}
and approximate the gradient by its leading term in $\gamma$
\begin{equation}
    ||\nabla x_0||_2\approx\frac{\gamma^{3L}}{\sqrt{2}}.
    \label{ezne_grad_3}
\end{equation}

Substitution of $||\nabla x_0||_2$ into the shot variance Eq.~\eqref{var_x0_3} gives
\begin{equation}
    \sigma^2=\Var(x_0)=\frac{3\gamma^{6L}}{2s},\label{var_x0_4}
\end{equation}
which we solve for $s$ to yield
\begin{equation}
    s=\frac{3\gamma^{6L}}{2\sigma^2}.\label{ezne_scaling}
\end{equation}
To generalize this result from a single Pauli string with unit weight to that of a general expectation value, as in the derivation of the sampling overhead scaling of RIDA in SM Sec.~\ref{appendix:scaling}, we apply the rule for the variance of a linear combination of random variables to incorporate a list of Pauli strings with different weights, which gives
\begin{equation}
    s=\frac{3||\boldsymbol{x}||^2}{2\sigma^2}\gamma^{6L},\label{ezne_scaling_2}
\end{equation}
where $\boldsymbol{x}$ is again a vector of weights. Thus, exponential ZNE has cubically larger overhead scaling when compared to RIDA, as supported by the observed shot error scaling in numerical tests provided in SM Sec.~\ref{appendix:predictions}.

\subsection{CNOT-only depolarization + quadratic ZNE overhead}
\label{appendix:qzne_overhead}

In order to analyze the sampling overhead of CNOT-only depolarization + quadratic ZNE, we first examine the sampling overhead of quadratic ZNE in isolation, which is equivalent to the total sampling overhead when $p\rightarrow0$. According to quadratic extrapolation of the data points $x_1$, $x_3$, and $x_5$ to the estimated error-free expectation value $x_0$  Eq.~\eqref{x0_qfit}, the variance of the error-free estimate of quadratic ZNE is
\begin{equation}
    \sigma^2=\Var(x_0)=\Var\left(\frac{3}{8}x_5-\frac{10}{8}x_3+\frac{15}{8}x_1\right).\label{qzne_overhead_1}
\end{equation}
We decompose this variance by recognizing that, since $x_1$, $x_3$, and $x_5$ represent individual circuits that are sampled independently, the variance may be expressed as
\begin{align}
    \sigma^2&=\Var\left(\frac{3}{8}x_5\right)+\Var\left(-\frac{10}{8}x_3\right)+\Var\left(\frac{15}{8}x_1\right)\\
    &=\left(\frac{3}{8}\right)^2\Var(x_5)+\left(\frac{10}{8}\right)^2\Var(x_3)+\left(\frac{15}{8}\right)^2\Var(x_1).\label{qzne_overhead_2}
\end{align}
and since in the zero-error limit where $p\rightarrow 0$, the variances of $x_1$, $x_3$, and $x_5$ are identically distributed, the variance may be expressed as
\begin{equation}
    \sigma^2=\frac{167}{32}\Var(x_1);\label{qzne_overhead_3}
\end{equation}
The shot variance Eq.~\eqref{var_o} of quadratic ZNE in the zero-error limit $\langle O_\epsilon\rangle=\langle O\rangle$ is thus 
\begin{equation}
    \sigma^2=\frac{167}{32}\left(\frac{1-\langle O\rangle^2}{s}\right),\label{qzne_overhead_4}
\end{equation}
such that the dependence of the shot number $s$ on the variance $\sigma^2$ is
\begin{equation}
    s=\frac{167}{32}\left(\frac{3-3\langle O\rangle^2}{\sigma^2}\right).\label{qzne_overhead_5}
\end{equation}
\\

It is worth noting that, in contrast, RIDA performs no mitigation in the zero-error limit, such that its variance can be readily equated with the unmitigated variance Eq.~\eqref{var_o} and solved to find
\begin{equation}
    s_0=\left(\frac{1-\langle O\rangle^2}{\sigma^2}\right),\label{RIDA_overhead_comp}
\end{equation}
where $s_0$ is the number of shots required from RIDA. Equation of the variance of quadratic ZNE Eq.~\eqref{qzne_overhead_5} and RIDA \eqref{RIDA_overhead_comp} thus yields
\begin{equation}
    s=\frac{501}{32}s_0\approx15.66s_0,
\end{equation}
such that the quadratic ZNE that underlies conventional use of CNOT-only depolarization \cite{PhysRevLett.127.270502} entails $15.66\times$ more shots than RIDA in the zero-error limit. Note that since quadratic ZNE circuits include $1\times$, $3\times$, and $5\times$ as many gates as the unmitigated circuit for the $\lambda=1,3,5$ circuits, respectively, quadratic ZNE in fact calls for $9\times$ more gates for the same number of shots per circuit (as well as $3\times$ more shots per circuit). Thus, a total of $46.97\times$ more gates are used in quadratic ZNE in the zero-error limit than RIDA.
\\

To expand upon the sampling overhead of quadratic ZNE to obtain the sampling overhead of CNOT-only depolarization + quadratic ZNE, we recall the equation for variance of each extrapolated data point (\textit{i.e.}, $x_1$, $x_3$, and $x_5$) in the high-error limit Eq.~\eqref{var_o_extrapolate_approx}. Given that the error-free expectation value for each $x_\lambda$ is a product of its noisy expectation value and depolarization probability factor $1/(1-\hat{p}_\lambda)$ (which leads to amplification of the corresponding variance by $1/(1-\hat{p}_\lambda)^2$), the variance of CNOT-only depolarization + quadratic ZNE in the high-error limit is
\begin{equation}
    \sigma^2=\frac{3}{s}\left(\frac{3}{8}\right)^2\left(\frac{1}{1-\hat{p}_5}\right)^{2}+\frac{3}{s}\left(\frac{10}{8}\right)^2\left(\frac{1}{1-\hat{p}_3}\right)^2+\frac{3}{s}\left(\frac{15}{8}\right)^2\left(\frac{1}{1-\hat{p}_1}\right)^2.\label{qzne_overhead_6}
\end{equation}
Although the estimated depolarization probabilities are not necessarily equal to the true depolarization probabilities $\hat{p}_\lambda\neq p_\lambda$, in order for CNOT-only depolarization to be effective, this estimate must be reasonably accurate. Thus, we make the assumption $\hat{p}_\lambda\approx p_\lambda$. Where we define $p:=p_1$, substitution of the relationships between the depolarization probabilities for $x_1$, $x_3$, and $x_5$ Eq.~\eqref{x3-x1_relationship} Eq.~\eqref{x5-x3_relationship} then yields
\begin{equation}
    \sigma^2=\frac{3}{s}\left(\frac{3}{8}\right)^2\left(\frac{1}{1-p}\right)^{10}+\frac{3}{s}\left(\frac{10}{8}\right)^2\left(\frac{1}{1-{p}}\right)^6+\frac{3}{s}\left(\frac{15}{8}\right)^2\left(\frac{1}{1-p}\right)^2,\label{qzne_overhead_7}
\end{equation}
such that expression in terms of the single-layer depolarization probability Eq.~\eqref{rida_pstar} and solution for $s$ gives
\begin{equation}
    s=\frac{3}{\sigma^2}\left(\frac{3}{8}\right)^2\left(\frac{1}{1-p^*}\right)^{10L}+\frac{3}{\sigma^2}\left(\frac{10}{8}\right)^2\left(\frac{1}{1-p^*}\right)^{6L}+\frac{3}{\sigma^2}\left(\frac{15}{8}\right)^2\left(\frac{1}{1-p^*}\right)^{2L},\label{qzne_overhead_8}
\end{equation}
which can in turn be expressed in terms of the error strength $\gamma$ Eq.~\eqref{gamma} as
\begin{equation}
    s=\frac{3}{\sigma^2}\left(\frac{3}{8}\right)^2\gamma^{10L}+\frac{3}{\sigma^2}\left(\frac{10}{8}\right)^2\gamma^{6L}+\frac{3}{\sigma^2}\left(\frac{15}{8}\right)^2\gamma^{2L},\label{qzne_overhead_9}
\end{equation}
of which the highest-order term gives
\begin{equation}
    s\approx\frac{27}{64\sigma^2}\gamma^{10L}\label{qzne_overhead_10}.
\end{equation}
Application of the rule for the variance of a linear combination of random variables to incorporate a list of Pauli strings, as discussed for RIDA (SM Sec.~\ref{appendix:scaling}), then gives the sampling overhead scaling of CNOT-only depolarization + quadratic ZNE
\begin{equation}
    s=\frac{27||\boldsymbol{x}||^2}{64\sigma^2}\gamma^{10L},\label{qzne_overhead_11}
\end{equation}
where $||\boldsymbol{x}||$ is again a vector of weights. Thus, CNOT-only depolarization + quadratic ZNE has quintically larger overhead scaling than RIDA where the depolarization probability is accurate. In practice, the scaling is instead quintic with respect to the estimated depolarization probability, which often leads to reduced overhead at the cost of inaccurate estimation of the error-free expectation; the quintic bound represents the required overhead in order to achieve high accuracy.

\section{Extended Results}

\subsection{Convergence of the depolarization probability estimate in RIDA}
\label{appendix:convergence}

\begin{figure}
    \includegraphics[scale=0.5]{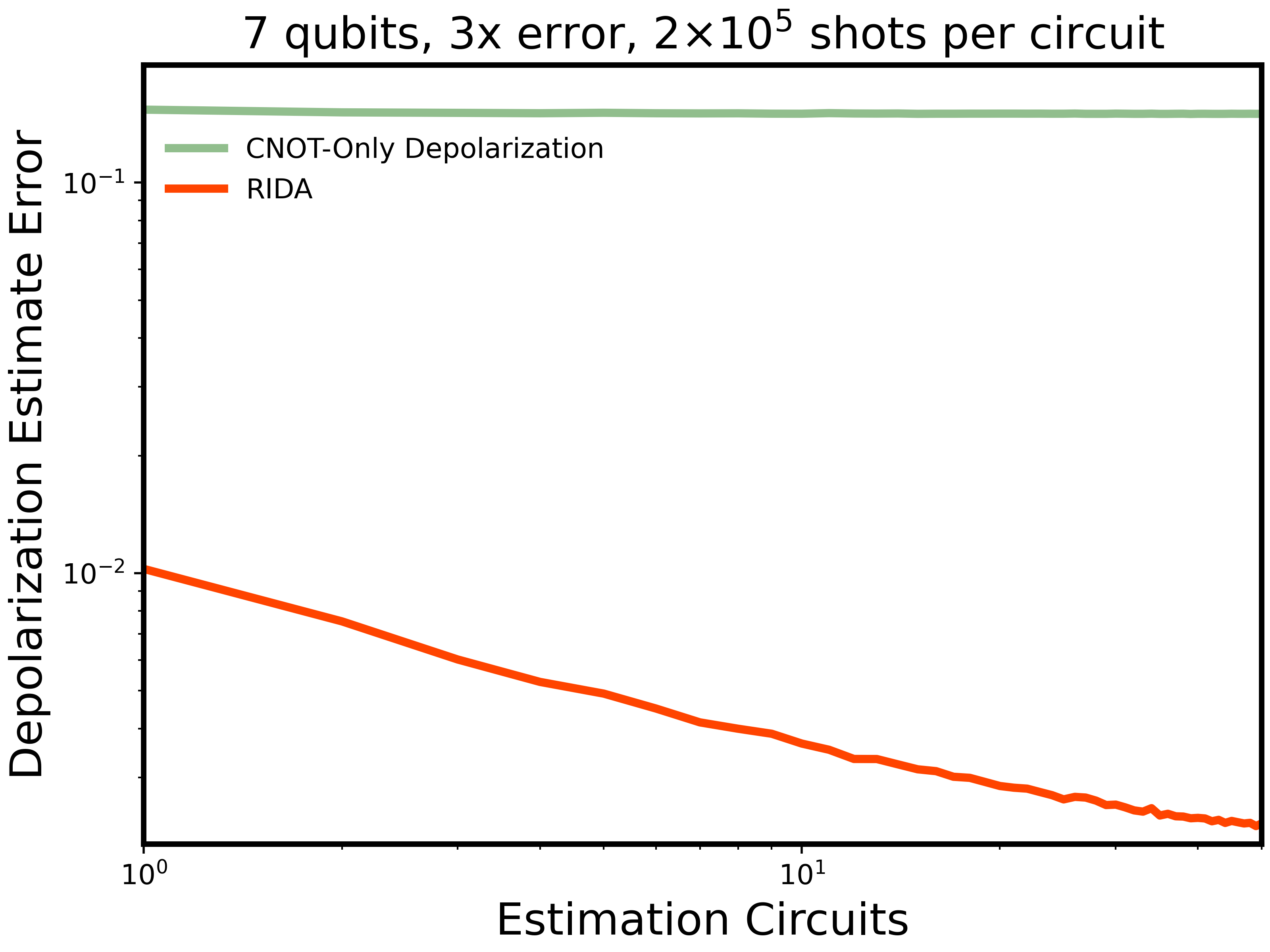}
    \caption{RMSE of the depolarization probability as a function of the number of RIDA (orange line) and CNOT-only depolarization (green line) estimation circuits (integers in the range $[1,50]$) as compared to the optimal depolarization probability estimate, which minimizes the least-squares error on trial data. 
    Individual data points represent an average across $5000$ random subsets of a pool of $p$-values obtained from $500$ circuits, an approach that reduces the computing time of numerical simulations from weeks to minutes (see code \cite{ridagithub}).
    \label{figure:convergence}}
\end{figure}

As shown in Fig.~\ref{figure:convergence}, the RMSE of the depolarization probability $p$ of RIDA is significantly lower than that of CNOT-only depolarization for all number of estimation circuits considered -- in the case of a single estimation circuit, by an order of magnitude. Furthermore, as the number of estimation circuits increases, the computed $p$ continues to become more accurate.

\subsection{CNOT-only depolarization with random rotations}\label{appendix:rotations}

\begin{figure}[ht]
    \begin{overpic}[scale=0.5]{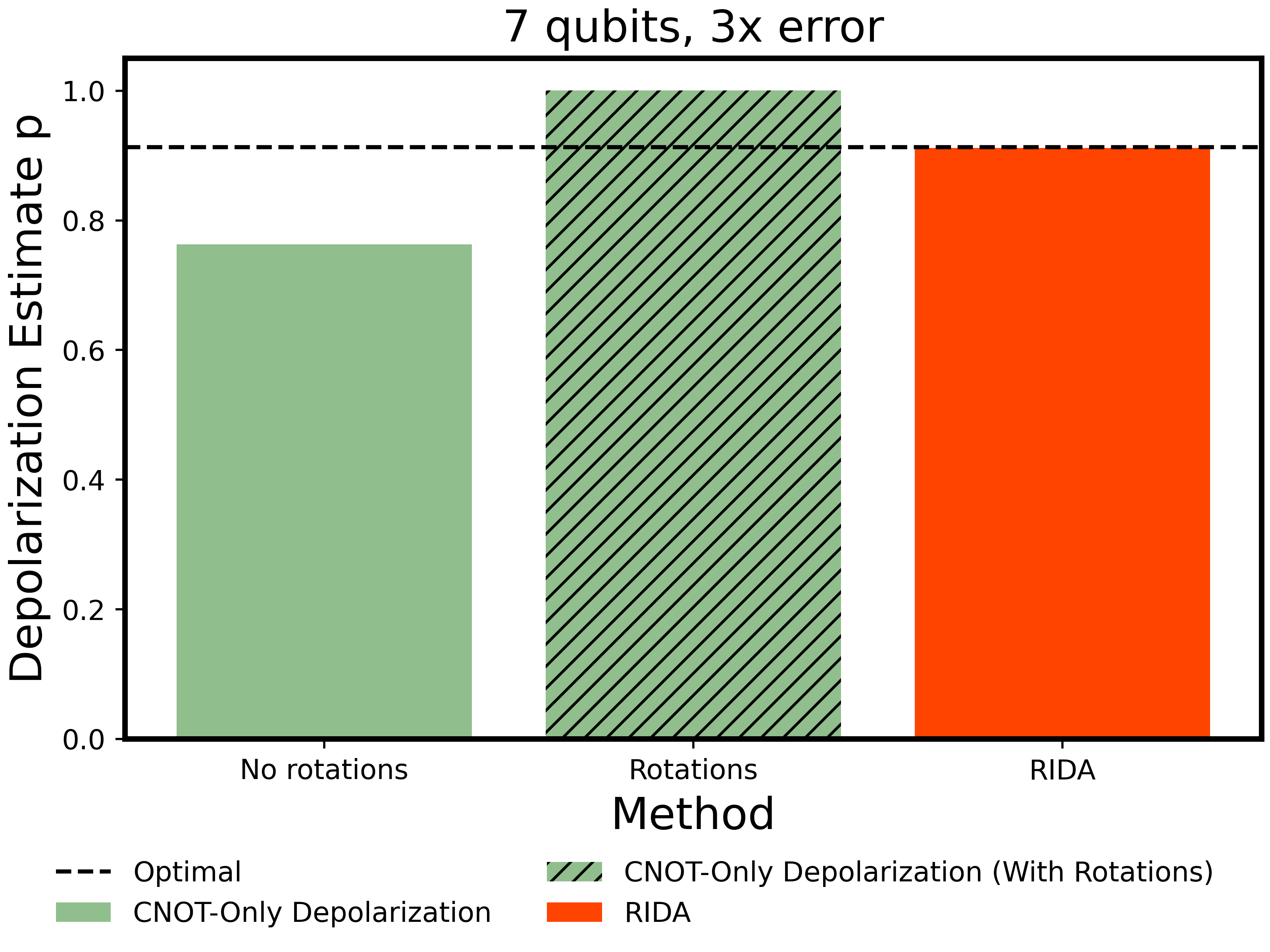}
    \put(3.5,70.5){\pf{(a)}}
    \end{overpic}
    \begin{overpic}[scale=0.5]{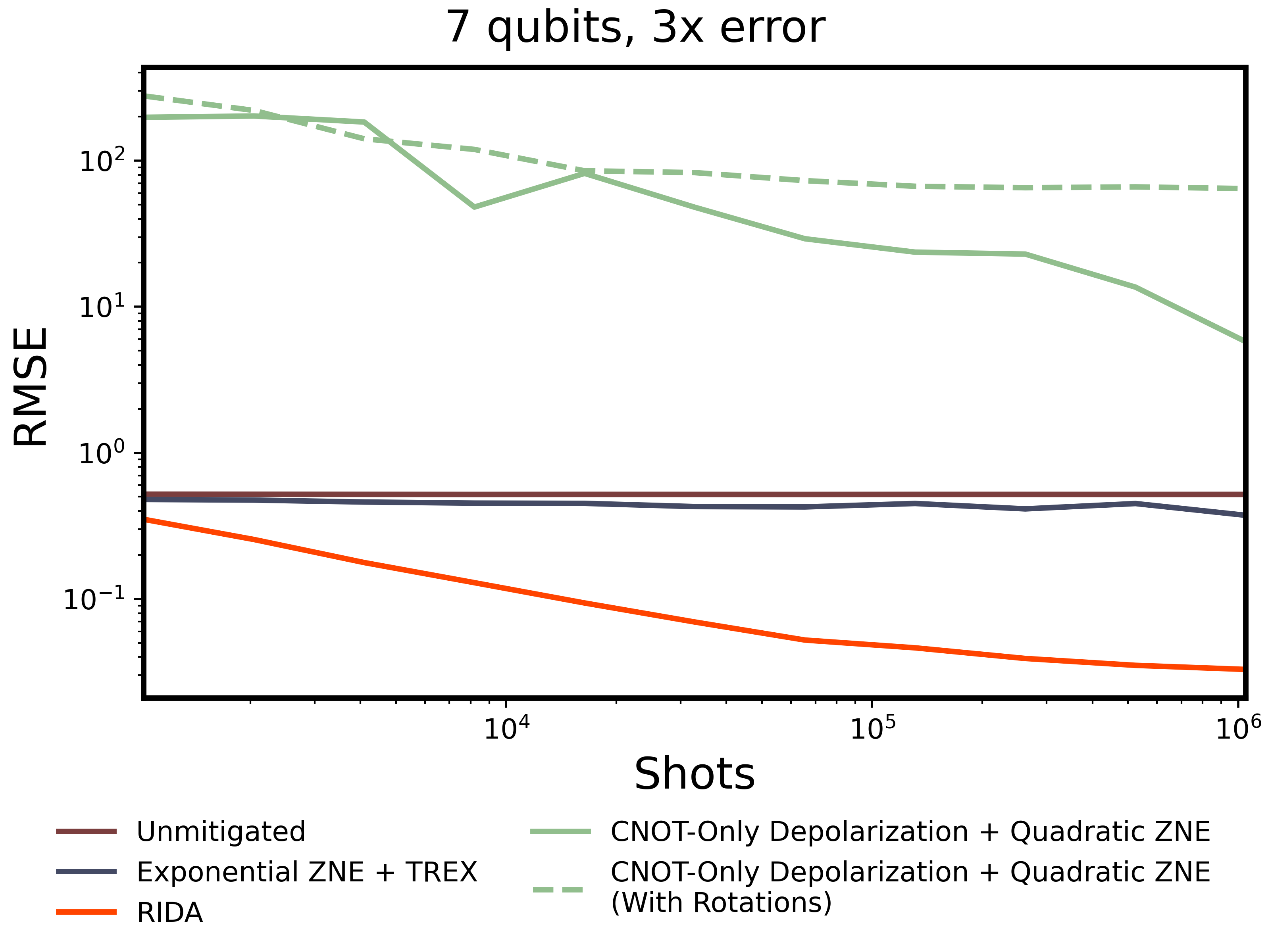}
    \put(3.5,70.5){\pf{(b)}}
    \end{overpic}
    \caption{(a) Magnitude of the depolarization probability estimate for CNOT-only depolarization with and without random rotation layers (green bars; solid and striped, respectively) and RIDA (orange bar) as compared to the optimal depolarization probability (dashed black line). The CNOT-only depolarization and RIDA estimates shown represent averages over the estimates for all Pauli strings considered, and the optimal depolarization probability shown follows from minimization of the RMSE on the trial data. (b) RMSE as a function of shot number for CNOT-only depolarization with quadratic ZNE both with and without random rotation layers (green lines; dashed and solid, respectively) as compared to RIDA (solid orange line), exponential ZNE + TREX (solid blue line), and the unmitigated results (solid brown line). }
    \label{figure:cnot_combined}
\end{figure}

Comparison of the depolarization probability estimate from CNOT-only depolarization with and without random rotation layers indicates the latter tends to be significantly more accurate independent of the error level, as exemplified by Fig.~\ref{figure:cnot_combined}[a]). For the EfficientSU2 ansatz considered here, with random rotation layers, the depolarization probability frequently approaches the singularity in the depolarizing model at $p=1$, a finding consistent with error-free circuit execution that does not result in a return to the initial state, as described in see SM Sec.~\ref{appendix:cnot}. Fig.~\ref{figure:cnot_combined} suggests the higher error in $p$ observed for CNOT-only depolarization with random rotation layers is associated with often significantly higher RMSE than alternative methods considered, including CNOT-only depolarization without random rotation layers.

Note that in the original formulation of CNOT-only depolarization \cite{PhysRevLett.127.270502}, the target circuit structure was such that random rotation layers could be used while maintaining an error-free expectation of $\langle O\rangle=1$. For such a system, the random rotation layers improved the accuracy of depolarization probability estimates and hence improved the accuracy of the CNOT-only depolarization method as a whole. However, the depolarization probability estimate nonetheless required quadratic ZNE as a supplemental method, and thus led to similar quintic scaling issues as described in SM Sec.~\ref{appendix:qzne_overhead}.

\subsection{Extended error multiplier results}
\label{appendix:multipliers}
\begin{figure*}[ht]
    \begin{overpic}[width=\textwidth]{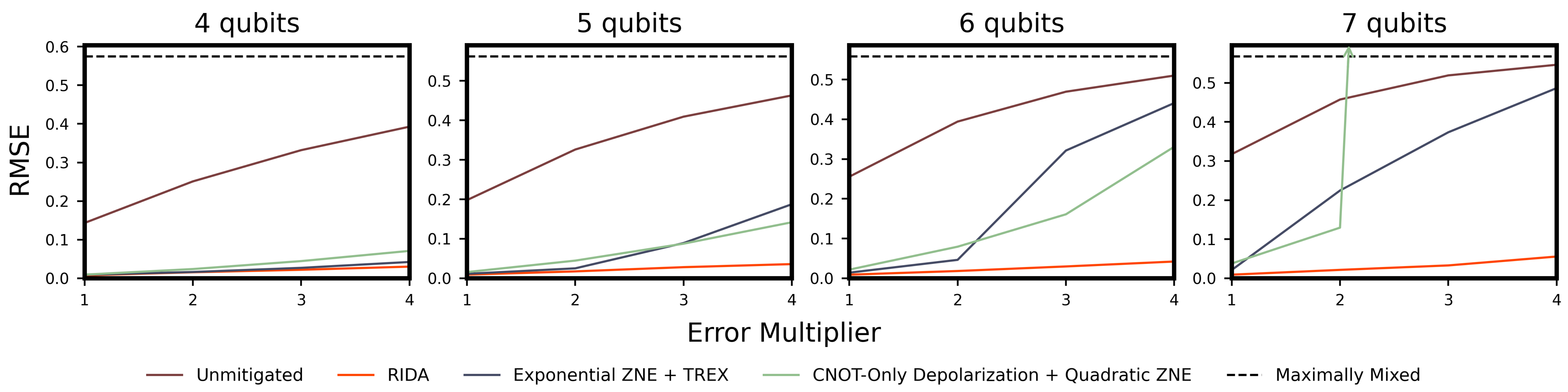}
    \put(0.2, 22){\pf{(a)}}
    \put(26.7, 22){\pf{(b)}}
    \put(51.1, 22){\pf{(c)}}
    \put(75.5, 22){\pf{(d)}}
    \end{overpic}
    \caption{RMSE as a function of the error multiplier for qubit numbers $\{4,5,6,7\}$ for RIDA (orange line), exponential ZNE + TREX (blue line), CNOT-only depolarization + quadratic ZNE (green line), and unmitigated results (brown line).}
    \label{figure:multipliers}
\end{figure*}

Fig.~\ref{figure:multipliers} expands upon the results in the main text that indicate that RIDA scales more favorably than exponential ZNE + TREX and CNOT-only depolarization + quadratic ZNE for seven qubits to show that the behavior persists for all qubit numbers considered and prevails to high error levels. 

\subsection{Extended shot number results}
\label{appendix:shot_results}
\begin{figure*}
    \begin{overpic}[width=\textwidth]{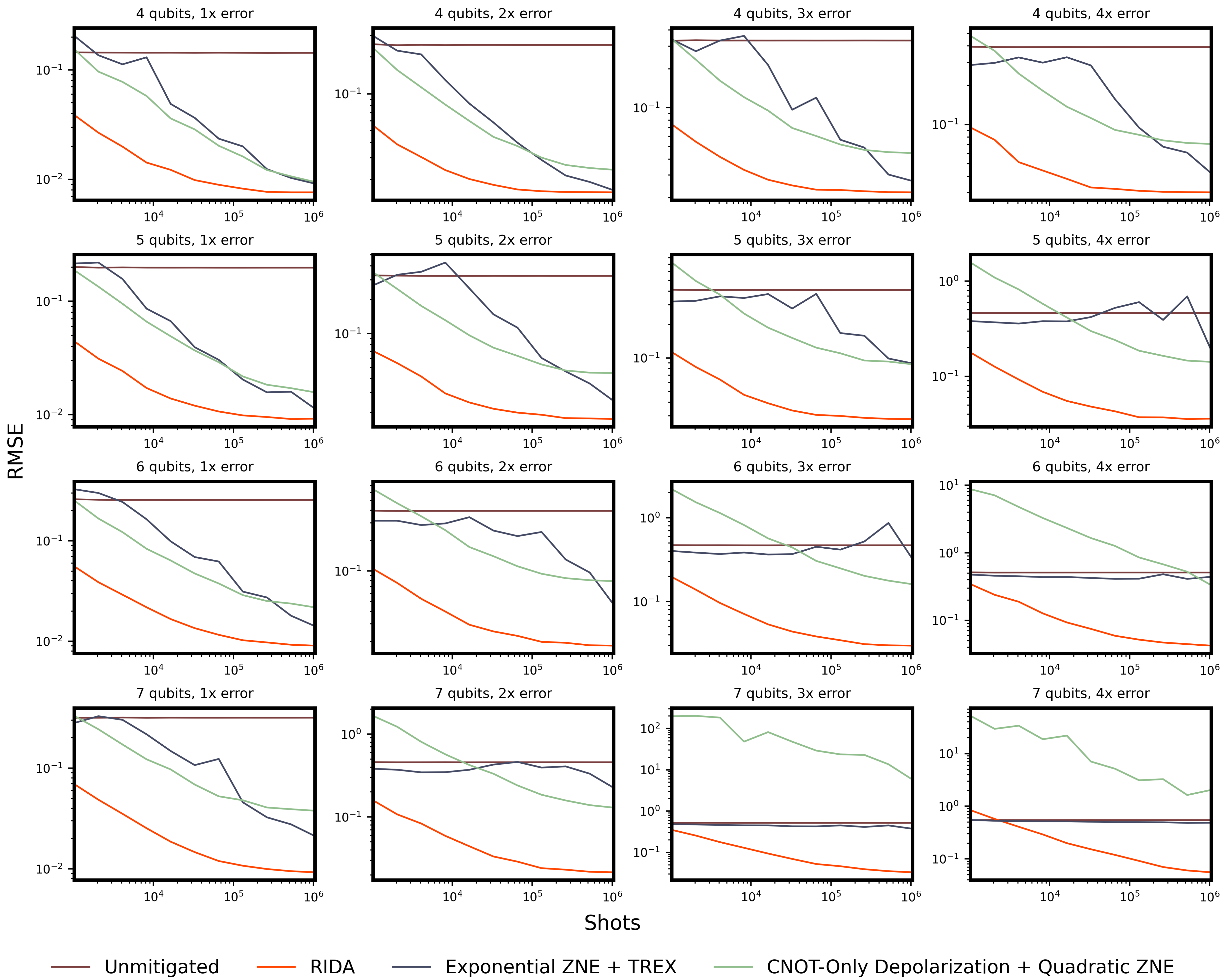}
    \put(2.5, 77){\pf{(a)}}
    \put(27, 77){\pf{(b)}}
    \put(51.5, 77){\pf{(c)}}
    \put(76, 77){\pf{(d)}}
    \put(2.5, 58.5){\pf{(e)}}
    \put(27, 58.5){\pf{(f)}}
    \put(51.5, 58.5){\pf{(g)}}
    \put(76, 58.5){\pf{(h)}}
    \put(2.5, 40){\pf{(i)}}
    \put(27, 40){\pf{(j)}}
    \put(51.5, 40){\pf{(k)}}
    \put(76, 41){\pf{(l)}}
    \put(2.5, 21.5){\pf{(m)}}
    \put(27, 21.5){\pf{(n)}}
    \put(51.5, 21.5){\pf{(o)}}
    \put(76, 21.5){\pf{(p)}}
    \end{overpic}
    \caption{RMSE as a function of shot number for RIDA (orange line), exponential ZNE + TREX (blue line), CNOT-only depolarization + quadratic ZNE (green line), and unmitigated results (brown line).}
    \label{figure:shots}
\end{figure*}

Fig.~\ref{figure:shots} indicates for a broader range of qubit number and error multiplier combinations that RIDA consistently yields the lowest error of all error mitigation methods considered, a finding which supports the argument that RIDA reliably outperforms both exponential ZNE + TREX and CNOT-only depolarization + quadratic ZNE for many different conditions, with potential to extend to larger quantum systems that have greater total error. Note that, although we show very high RMSE for CNOT-only depolarization in the high-error limit, it is not necessarily inferior to exponential ZNE. For small amounts of error and moderate shot numbers, similar to the tests in the original presentation of the method \cite{PhysRevLett.127.270502}, CNOT-only depolarization yields modest improvement over exponential ZNE. Moreover, in the high-error limit, the difference in RMSE between CNOT-only depolarization and exponential ZNE (which defaults to little-to-no extrapolation) is not of practical importance since both are at least similar to the unmitigated RMSE; in this regime, of the three methods, only RIDA is viable.

\subsection{Shot error prediction results}
\label{appendix:predictions}
\begin{figure*}
    \begin{overpic}[width=\textwidth]{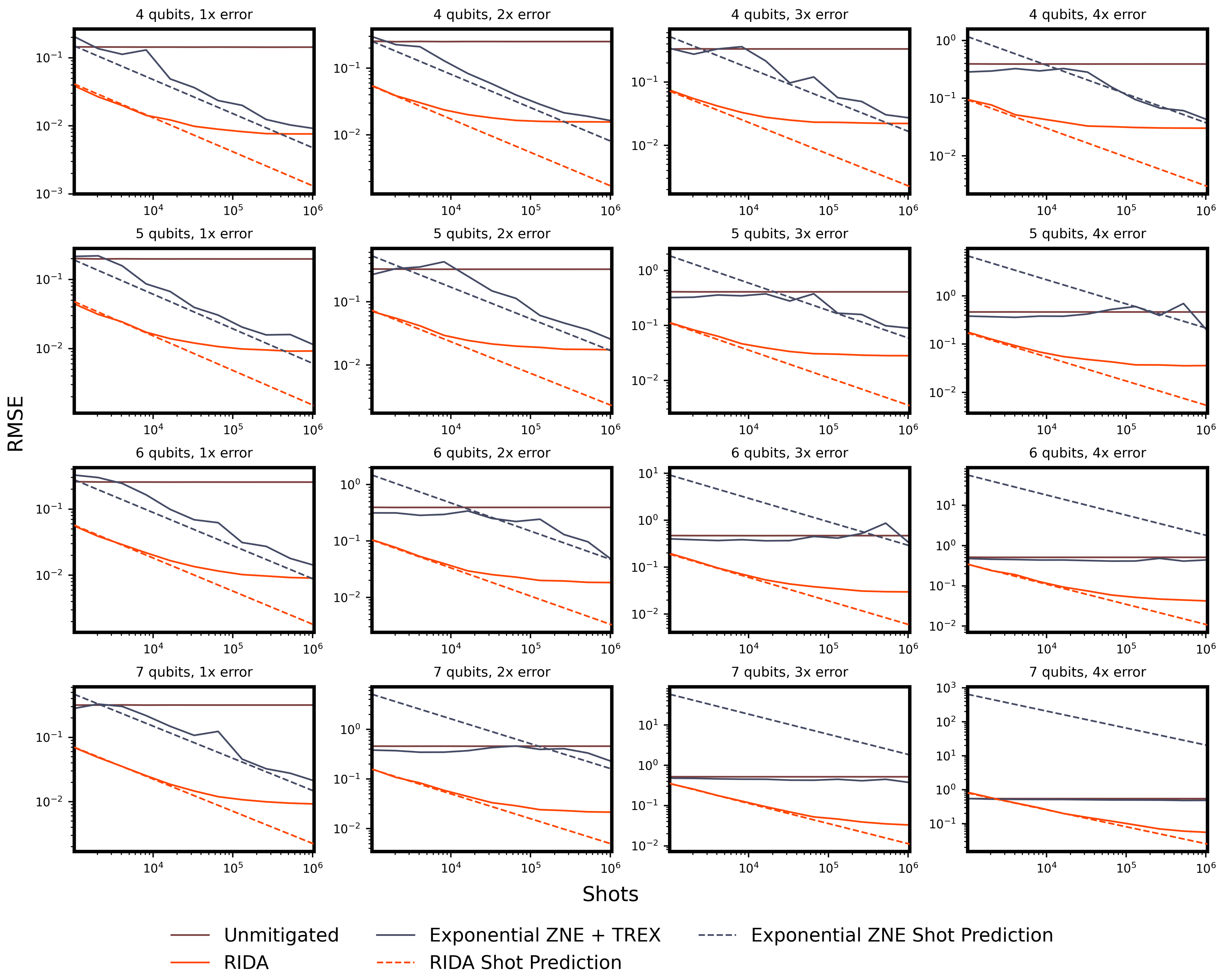}
    \put(2.5, 77.5){\pf{(a)}}
    \put(27, 77.5){\pf{(b)}}
    \put(51.5, 77.5){\pf{(c)}}
    \put(76, 77.5){\pf{(d)}}
    \put(2.5, 59.5){\pf{(e)}}
    \put(27, 59.5){\pf{(f)}}
    \put(51.5, 59.5){\pf{(g)}}
    \put(76, 59.5){\pf{(h)}}
    \put(2.5, 41.5){\pf{(i)}}
    \put(27, 41.5){\pf{(j)}}
    \put(51.5, 42.5){\pf{(k)}}
    \put(76, 41.5){\pf{(l)}}
    \put(2.5, 23.5){\pf{(m)}}
    \put(27, 23.5){\pf{(n)}}
    \put(51.5, 23.5){\pf{(o)}}
    \put(76, 25){\pf{(p)}}
    \end{overpic}
    \caption{Comparison of the analytically predicted (dashed lines) and numerically obtained (solid lines) RMSE as a function of the shot number for RIDA (orange) and exponential ZNE (blue, with TREX for numerical results), with unmitigated result for comparison (brown line).}
    \label{figure:predictions}
\end{figure*}

Fig.~\ref{figure:predictions} shows that the analytically predicted RIDA and exponential ZNE shot variances [Eqs.~\eqref{rida_shots_2} and \eqref{var_x0_3}, respectively] align well with the corresponding numerically obtained RMSE values (see SM Sec.~\ref{appendix:scaling} and Sec.~\ref{appendix:zne_scaling}, respectively).
RIDA almost exactly matches its predicted shot error in the low-shot limit and deviates only for larger shot numbers where nonshot error is expected to dominate. The behavior of exponential ZNE is somewhat distinct given its greater degree of error: At low-error levels, exponential ZNE likewise agrees with its predicted shot error in the low-shot limit. However, where the shot error is predicted to be greater than the error of the unmitigated method, the exponential ZNE error oscillates about that of the unmitigated method, as expected when exponential ZNE becomes unstable and defaults to little-to-no extrapolation. In these cases, the RMSE of exponential ZNE resumes its descent once the shot number increases to the point that the predicted shot error intersects with the unmitigated method error. Note this effect, combined with the analytic scaling predictions for RIDA and exponential ZNE, implies that in the high-error limit, RIDA requires significantly fewer shots to outperform the unmitigated method than exponential ZNE. As a stark example, for the highest qubit number ($7$) and error multiplier ($4\times$) shown, the aforementioned behavior suggests that to improve upon the RMSE of the unmitigated error, RIDA requires just $2000$ shots but exponential ZNE would require approximately $1.5$ billion shots to reliably do so.

\subsection{Extended depolarizing model results}
\label{appendix:depolarizations}

\begin{figure*}
    \begin{overpic}[width=\textwidth]{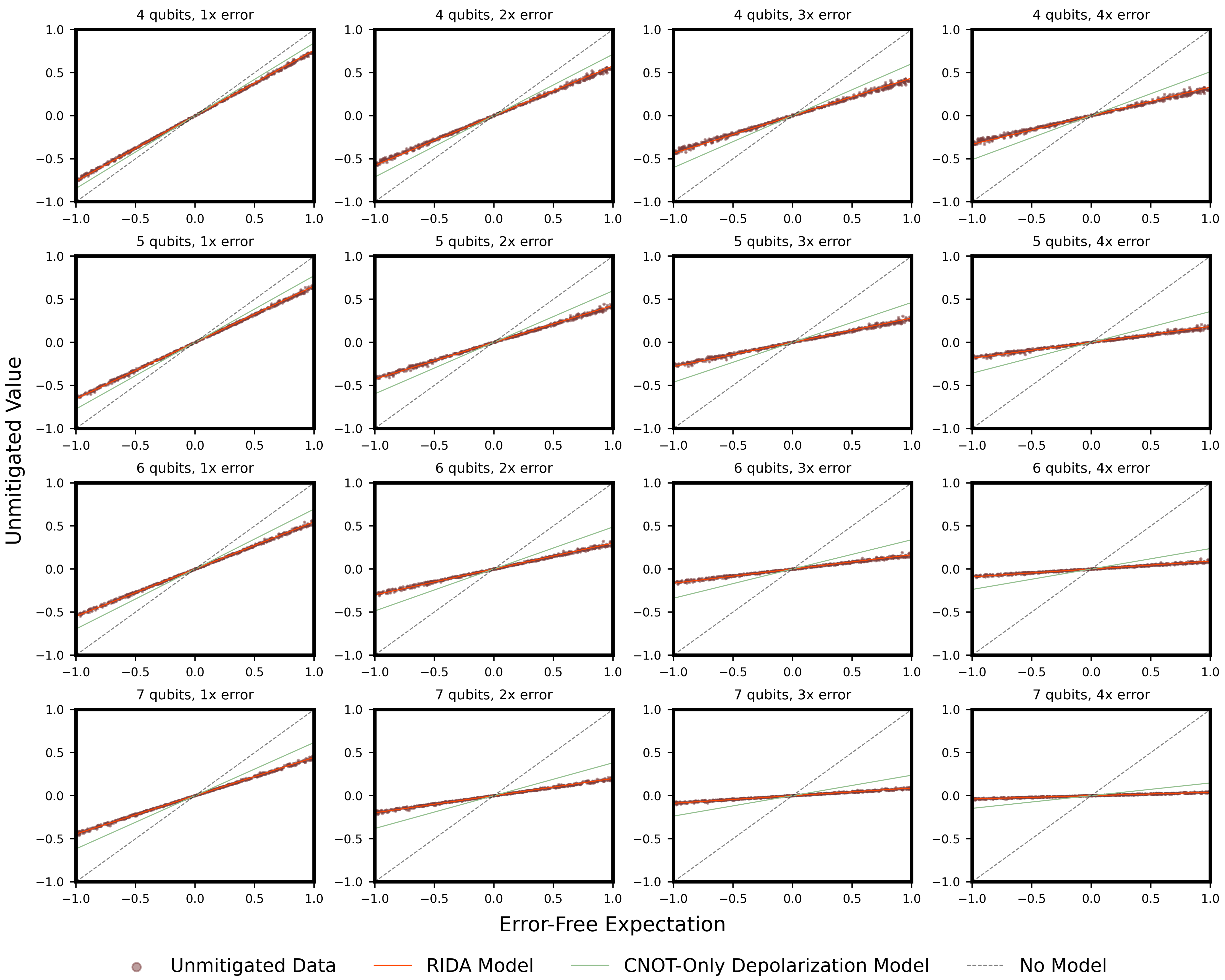}
    \put(4, 78.5){\pf{(a)}}
    \put(28.5, 78.5){\pf{(b)}}
    \put(53, 78.5){\pf{(c)}}
    \put(77.5, 78.5){\pf{(d)}}
    \put(4, 60){\pf{(e)}}
    \put(28.5, 60){\pf{(f)}}
    \put(53, 60){\pf{(g)}}
    \put(77.5, 60){\pf{(h)}}
    \put(4, 41.5){\pf{(i)}}
    \put(28.5, 41.5){\pf{(j)}}
    \put(53, 41.5){\pf{(k)}}
    \put(77.5, 41.5){\pf{(l)}}
    \put(4, 23){\pf{(m)}}
    \put(28.5, 23){\pf{(n)}}
    \put(53, 23){\pf{(o)}}
    \put(77.5, 23){\pf{(p)}}
    \end{overpic}
    \caption{Predicted relationship between the error-free and unmitigated expectation values according to the RIDA and CNOT-only depolarization models (solid lines; orange and green, respectively) as compared to unmitigated data (brown circles) and the result in the absence of a model (dashed gray line) for a variety of qubit number and error multiplier combinations. Depolarization probability estimates shown represent averages over all Pauli strings considered.}
    \label{figure:depolarization}
\end{figure*}

\begin{figure*}
    \begin{overpic}[width=\textwidth]{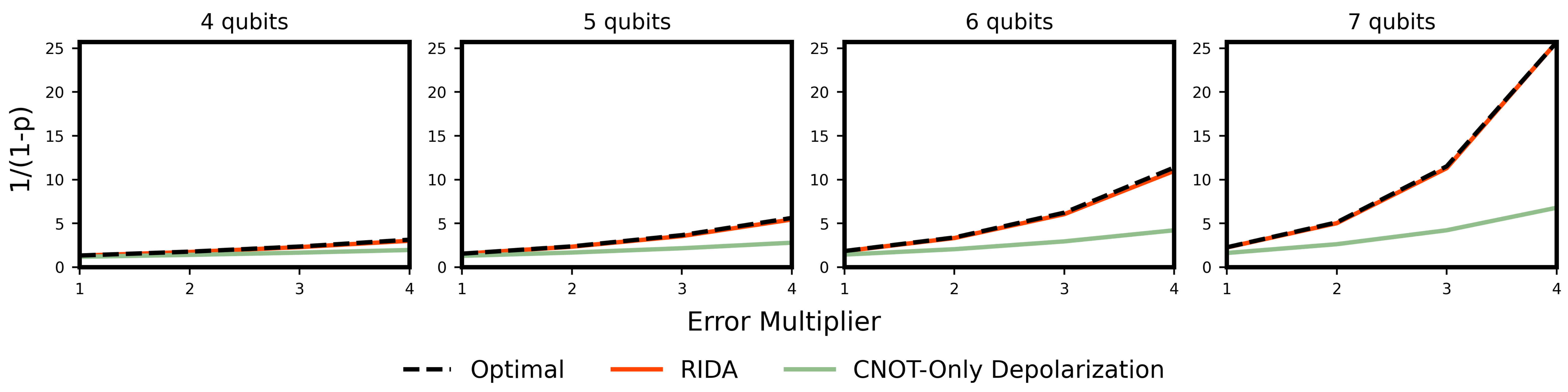}
    \put(3.3, 23.2){\pf{(a)}}
    \put(27.7, 23.2){\pf{(b)}}
    \put(52.1, 23.2){\pf{(c)}}
    \put(76.5, 23.2){\pf{(d)}}
    \end{overpic}
    \caption{Estimated error strength $1/(1-p)$ as a function of the error multiplier for qubit numbers $\{4,5,6,7\}$ according to RIDA and CNOT-only depolarization model results averaged over all Pauli strings (solid lines; orange and green, respectively) and the optimal fit that minimizes the RMSE for the trial data (dashed black line).}
    \label{figure:depolarization_summary}
\end{figure*}

The ability of RIDA to more accurately approximate the depolarization probability than CNOT-only depolarization is further supported by Figs.~\ref{figure:depolarization} and \ref{figure:depolarization_summary}.
Fig.~\ref{figure:depolarization} demonstrates that the more accurate approximation of the depolarization probability than CNOT-only depolarization for $7$ qubits and $3\times$ error illustrated in the main text holds for all combinations of qubit number and error multiplier considered. Furthermore, Fig.~\ref{figure:depolarization_summary} reveals that in all cases considered, the optimal and RIDA estimates of the depolarization error strength $\frac{1}{1-p}$ [the constant factor applied to the noisy expectation value to yield the error-free expectation in the depolarizing model Eq.~\eqref{true_expectation}] are indistinguishable to the eye, even as the difference between optimal and CNOT-only depolarization estimates monotonically increases with the qubit number and error multiplier.

\subsection{Extended Pauli string error results}
\label{appendix:paulis}
\begin{figure*}
    \begin{overpic}[width=\textwidth]{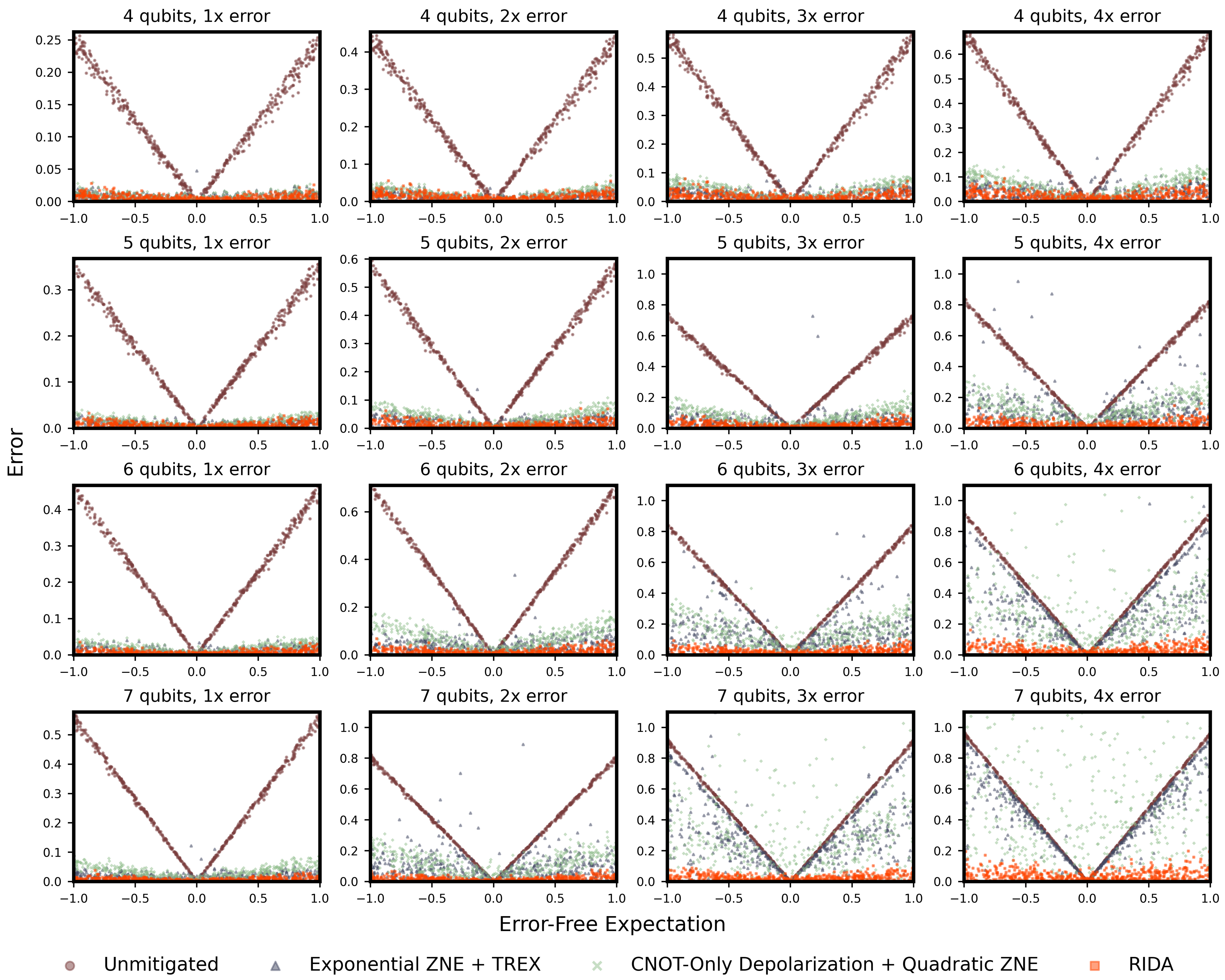}
    \put(4, 78.5){\pf{(a)}}
    \put(28.5, 78.5){\pf{(b)}}
    \put(53, 78.5){\pf{(c)}}
    \put(77.5, 78.5){\pf{(d)}}
    \put(4, 60){\pf{(e)}}
    \put(28.5, 60){\pf{(f)}}
    \put(53, 60){\pf{(g)}}
    \put(77.5, 60){\pf{(h)}}
    \put(4, 41.5){\pf{(i)}}
    \put(28.5, 41.5){\pf{(j)}}
    \put(53, 41.5){\pf{(k)}}
    \put(77.5, 41.5){\pf{(l)}}
    \put(4, 23){\pf{(m)}}
    \put(28.5, 23){\pf{(n)}}
    \put(53, 23){\pf{(o)}}
    \put(77.5, 23){\pf{(p)}}
    \end{overpic}
    \caption{Error as a function of the error-free expectation value across qubit numbers and error multipliers in the incoherent error model for RIDA (orange squares), exponential ZNE + TREX (blue triangles), CNOT-only depolarization + quadratic ZNE (green x's), and unmitigated values (brown circles). Values beyond the plotted regions are truncated.}
    \label{figure:paulis}
\end{figure*}

\begin{figure*}
    \begin{overpic}[width=\textwidth]{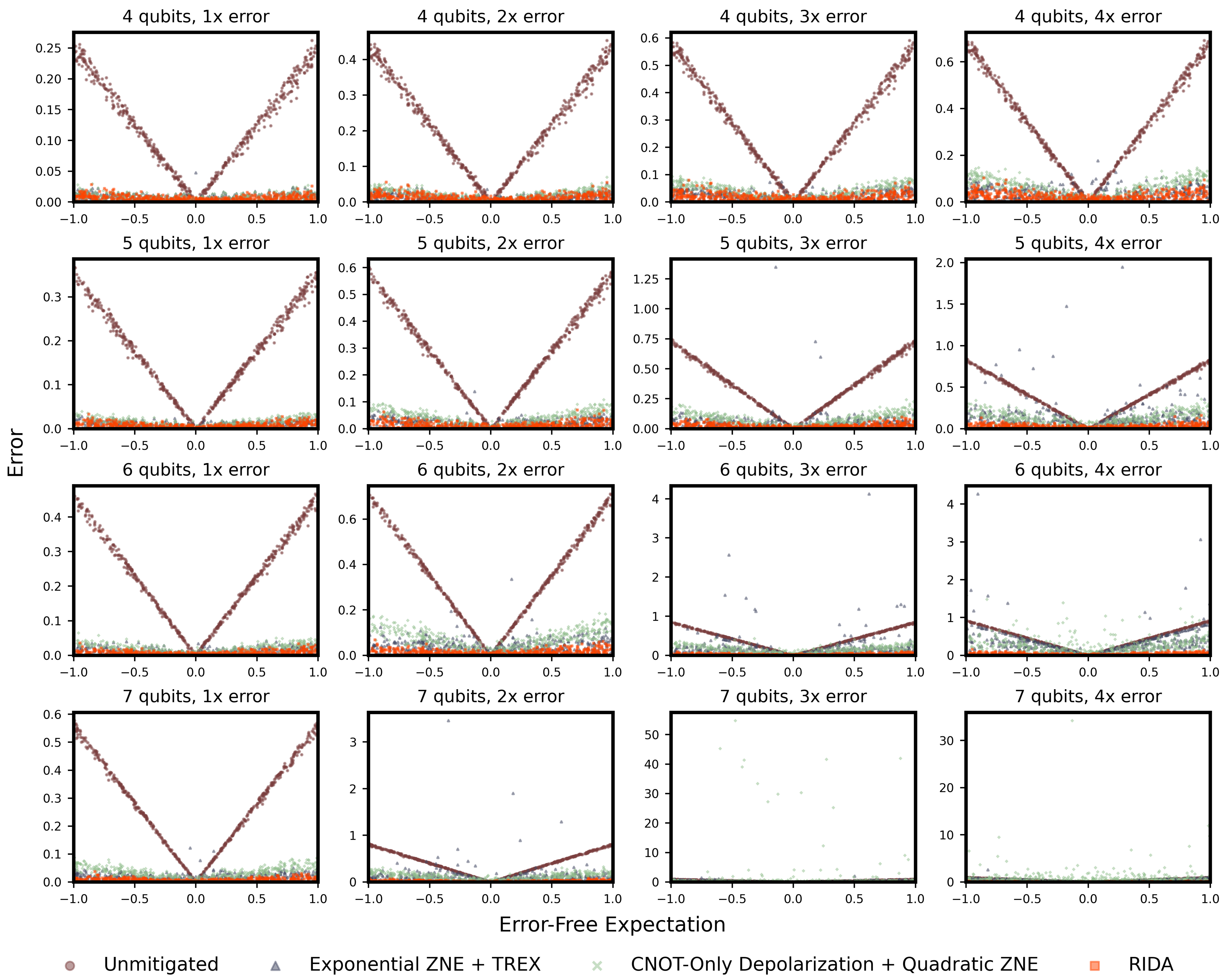}
    \put(4, 78.5){\pf{(a)}}
    \put(28.5, 78.5){\pf{(b)}}
    \put(53, 78.5){\pf{(c)}}
    \put(77.5, 78.5){\pf{(d)}}
    \put(4, 60){\pf{(e)}}
    \put(28.5, 60){\pf{(f)}}
    \put(53, 60){\pf{(g)}}
    \put(77.5, 60){\pf{(h)}}
    \put(4, 41.5){\pf{(i)}}
    \put(28.5, 41.5){\pf{(j)}}
    \put(53, 41.5){\pf{(k)}}
    \put(77.5, 41.5){\pf{(l)}}
    \put(4, 23){\pf{(m)}}
    \put(28.5, 23){\pf{(n)}}
    \put(53, 23){\pf{(o)}}
    \put(77.5, 23){\pf{(p)}}
    \end{overpic}
    \caption{Same as Fig.~\ref{figure:paulis} without truncation.}
    \label{figure:not_truncated}
\end{figure*}

Fig.~\ref{figure:paulis} illustrates that the behavior of the error as a function of the error-free expectation value illustrated for $7$ qubits, $3\times$ error, and an incoherent error model in the main text holds for all qubit numbers and error multipliers considered. Even as the RMSE of exponential ZNE + TREX and CNOT-only depolarization + quadratic ZNE rapidly increases at extremal error-free expectation values as the qubit number and error multiplier increases,\footnote{It is interesting to note that the unmitigated, exponential ZNE + TREX, and CNOT-only depolarization + quadratic ZNE results exhibit to differing extents a prominent ``V" shape. The V shape of the unmitigated results is consistent with the global depolarizing approximation in the unmitigated method, the discrepancy between the estimated and optimal depolarization probability for CNOT-only depolarization + quadratic ZNE, and the fact that implementation of exponential ZNE defaults to linear or no extrapolation where $x_1$, $x_3$, and $x_5$ yield undefined results (namely, where the shot error becomes dominant such that the probability $x_\lambda$ is a monotonic function becomes small, see SM Sec.~\ref{appendix:ezne}).}
the RMSE of RIDA remains persistently low throughout in contrast. Fig.~\ref{figure:not_truncated} additionally demonstrates that both exponential ZNE and CNOT-only depolariation + quadratic ZNE yield outliers with very large error as the overall error rate rises.

\bibliography{references}